# Tests of General Relativity: A Review

By Estelle Asmodelle

May 4th, 2017

**Dissertation for B.Sc. (Hons.) AA3050 in Astronomy, VSASTR513**

**University of Central Lancashire (UCLAN)**

# Abstract

This report is a literature review of significant and successful tests of general relativity [GR]. The GR predicted value for the perihelion advance of Mercury was $\Delta\phi$ = 43".03 century$^{-1}$ and fit well with observation, being the first success of GR. The GR result for the bending of light around the Sun $\delta_{GR}$ = 1".75, confirmed by observation, marked the second successful validation of GR. Gravitational Redshift [GvR] was first detected 1925 and is the third successful classical test. The parametrized post-Newtonian [PPN] employs $\beta$ and $\gamma$ for GR testing. Shapiro delay was also confirmed with $\gamma = 1.000021 \pm 0.000023$, against GR value $\gamma = 1$, some consider this to be the fourth classical test. So too gravitational time dilation [GvT] was experimentally confirmed in 1971, while GvT for GPS is a daily validation of GR. Frame−dragging and the Geodetic effect have also been confirmed. The strong equivalence principle [SEP] has been confirmed to $\eta = 4.4 \times 10^{-4}$, to GR $\eta = 0$. Gravitational slip has been constrained to $E_G = 0.48 \pm 0.10$ at $z = 0.32$, against the GR value, $E_G = 0.30 \pm 0.07$. The first gravitational wave detection from GW150914, in 2015, has confirmed a long-awaited phenomenon that has taken GR testing to higher precision. Gravitational lensing has also confirmed GR to better than 1%. GR continues to be tested, eliminating competing gravity theories.

*"Spacetime tells matter how to move, and matter tells spacetime how to curve,"*

John Wheeler, (Misner et al., 1973).

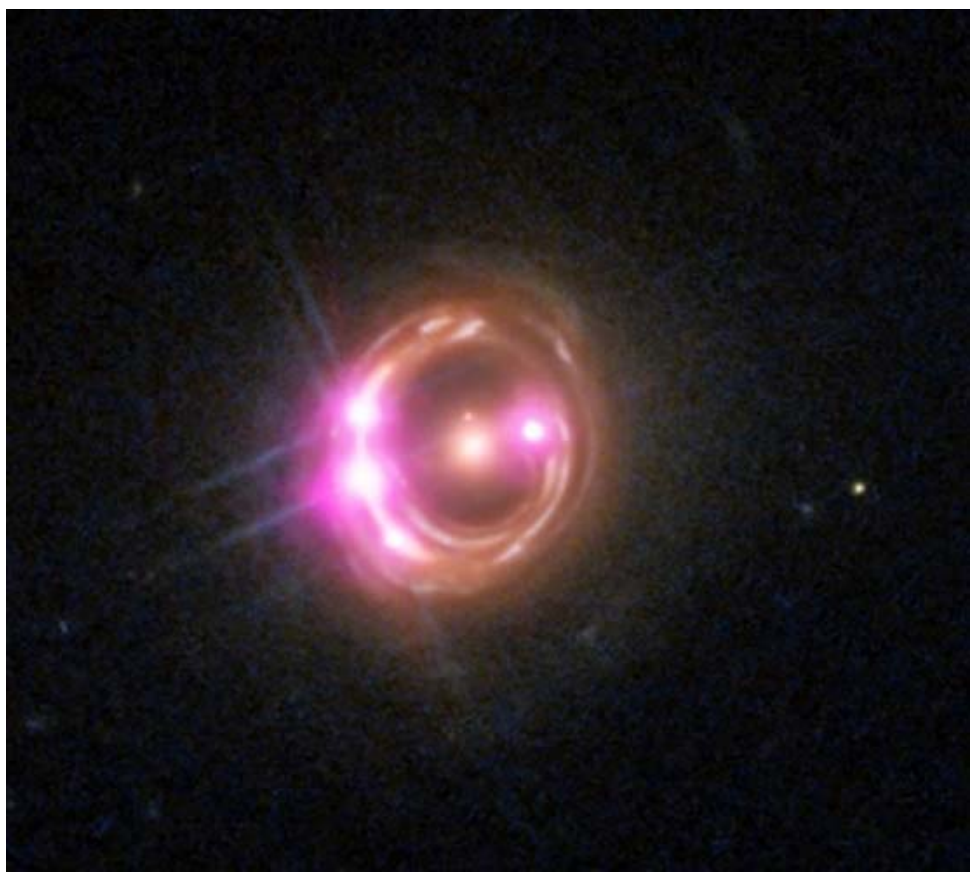

*Figure 1: General relativistic effects are profound. Composite image of quasar RX J1131-1231 lensed by a foreground elliptical galaxy. Image courtesy: X-ray: NASA/CXC/Univ of Michigan/R.C.Reis et al.; Optical: NASA/STScI.*

# CONTENTS

# Introduction

Since 1915 Einstein's general theory of relativity, over a hundred years, has remained unaltered and is fundamental to astrophysics and cosmology. General relativity and Einstein's field equations are considered by many as the epitome of physical law (Bean et al., 2011), and general relativity continues to be tested.

This report is a review of the first tests of the general theory of relativity and the core physical principles behind the observations and experimental arrangements. However, these tests are astronomical in nature, in that astronomical observational tests are evaluated, while *most* terrestrial experiments are outside the area of this report. The term 'first tests,' includes modern tests as well, and means the first test of each class. Although the first results are mentioned and compared to contemporary results, many subsequent results are not discussed.

Furthermore, the general theory of relativity is the currently accepted gravitational theory and as such a large repository of test results has been carried out since its inception in 1915. However, this report will only focus on what are considered the main tests, as there is no room to consider ancillary tests that have arisen. Needless to say, there is some historical context of the development of ideas and tests, yet the report is non-chronological. Moreover, the full mathematical description of general relativity is outside the field of this report. Many acronyms used in this report and a summary of their meanings are listed in Appendix A.

A basic introduction to the general theory of relativity is provided in Chapter 1, which introduces the main concepts of such tests. The physics behind the first of each of the evaluated test is examined, to some level, together with the mathematical foundation of the test and its relationship to general relativity. Most of the tests were carried out using the Schwarzschild metric. However, tests related to precession and frame dragging utilize the Kerr metric. Chapter 2, summarises the modern tests, while Chapter 3 examines briefly the strong−field tests. Chapter 4 glimpses at the results of cosmological tests.

Involved mathematical expressions and concepts, related to the report material, is outlined in a series of appendices for readers who wish to explore the material in more depth.

The purpose of this review is simply to investigate the validation of the general theory of relativity with contrasting modern data. Currently, in physics, astronomy, and cosmology there are several mysteries actively being pursued; namely, dark matter, dark energy, interiors of black holes, and the compatibility of general relativity with quantum mechanics. As such, many researchers are exploring the possibilities of different gravitational models to replace general relativity or methods for modifying it, see Appendix B for details.

However, Einstein's general theory of relativity is one of the most tested theories in physics and has never failed a single test. This report illustrates that success with results of crucial tests of the theory's validity.

# Chapter 1:

## Classical tests

### 1.1 The General Theory of Relativity

After Einstein published his special theory of relativity in 1905 (Einstein, 1905), the next step was to generalise his theory to include non-inertial reference frames; that is, to include acceleration and gravity. Later, Einstein published a paper in which he was in the first stages of generalising his theory of relativity to gravity (Einstein, 1907). As such, a few consequences of gravity were explored, including the bending of light, gravitational time dilation, and gravitational redshift. In that, he expounds his principle of equivalence[1], Einstein's equivalence principle [EEP]: that gravity and uniform acceleration are equivalent, locally (Einstein, 1907).

The static gravitational potential Φ, outside of a solid mass, using a Newtonian approximation is represented by Equation 1. This relation depicts the potential of a mass *M*, with radius *r.* See Appendix 1.1.1, for more details on the Newtonian gravitational potential.

$$\Phi = -\frac{GM}{r} \qquad \text{Eq. 1}$$

Based purely on EEP and the constancy of the speed of light, Einstein derived equations for both gravitational time dilation and gravitational redshift, to the first order. This work was several years before he had developed the correct version of the general theory of relativity.

Einstein showed that clocks in different gravitational potentials ran asynchronously, and derived gravitational time dilation for a homogeneous gravitational field[2] to the first order (Einstein, 1907). He started by having two clocks at two different positions in a given gravitational field, $\Phi_A$ and $\Phi_B$, whereby $\Phi_A - \Phi_B = gh$. The acceleration due to gravity is *g,* and the height from gravitating mass is *h*. The proper times at point A and B in that potential are then, $\Delta t_A$ and $\Delta t_B$, producing Equation 2. This relationship does not depend on the nature of the clocks and is a direct result of EEP.

$$\Delta t_B \simeq \left(1 + \frac{\Phi_A - \Phi_B}{c^2}\right) \Delta t_A \qquad \text{Eq. 2}$$

[1] The Galilean principle of equivalence or weak equivalence principle [WEP], rests on the equality of gravitational and inertial mass. While Einstein's equivalence principle [EEP] is concerned with the equality of homogeneous gravitation fields and acceleration.

[2] Einstein also made a footnote, stating that Equation 2 is also valid for an inhomogeneous gravitational field.

Hence, gravitational time dilation can be defined using Equation 3, where $t(0)$ is the coordinate time[3], $t(\Phi)$ is the local proper time[4] which runs slower than the coordinate time, and $\Phi$ is the gravitational potential. In Equation 3, $\Phi$ is always negative as seen from Equation 1. Clearly, $t(\Phi)$ is always less than $t(0)$, unless $\Phi = 0$. In such case, they are equal (Kenyon, 1991).

$$t(\Phi) \simeq t(0)\left[1+\frac{\Phi}{c^2}\right] \quad \text{Eq. 3}$$

In 1911, Einstein formulated the gravitational redshift in the same manner and derived its natural form (Einstein, 1911a), as in Equation 4, where $\nu_0$ is the emitting frequency of the rest frame of the source, and $\nu$ is the detected frequency. Einstein saw the gravitational redshift from EEP as a consequence of the first order Doppler shift, between the frames of A and B. In this formula, we see that the emitted frequency, or laboratory frequency, is dilated by the factor $[1 + \Phi/c^2]$, resulting in a lower observed frequency i.e. redshifted, to the first order.

$$\nu \simeq \nu_0\left[1+\frac{\Phi}{c^2}\right] \quad \text{Eq. 4}$$

Einstein assumed the gravitational potential started from the surface of a body and that the gravitational field was homogeneous and *static*. This relation was rearranged to provide a spectral shift in the following form, shown in Equation 5.

$$\frac{\nu_0-\nu}{\nu_0} \simeq \frac{-\Phi}{c^2} \quad \text{Eq. 5}$$

Einstein maintained the EEP and also Mach's principle[5], and explored a non-Euclidean geometric solution to a generalised theory, in curved four-dimensional spacetime. On the 25th November 1915, he finally published the correct version of the general theory of relativity [GR] (Einstein, 1915a). See Appendix 1.1.2 for more details of the field equations of GR.

Implementing Equations 1 and 2, now Equation 6 can be stated:

$$\mathrm{d}\tau^2 \simeq \mathrm{d}t^2\left(1+\frac{\Phi_A-\Phi_B}{c^2}\right) \simeq \mathrm{d}t^2\left(1+\frac{2\Phi}{c^2}\right) \simeq \mathrm{d}t^2\left(1-\frac{2GM}{rc^2}\right) \quad \text{Eq. 6}$$

where $\mathrm{d}\tau$ is the interval of proper time, and $\mathrm{d}t$ is the interval of coordinate time (Kenyon, 1991), while $\Phi_A$ is the first gravitational potential and $\Phi_B$ is the second gravitational potential. The

[3] The 'coordinate time' $t$, is measured by clocks that are far from gravity or acceleration (Schutz 2011).

[4] Hendrik Lorentz introduced 'local time,' which is the time of a clock that is stationary in a reference frame, such as in the theory of special relativity. In general relativity 'proper time' measures a clock that moves along a world lines, or the time between two events in spacetime. The local proper time $\tau$ is the time as measured by a local experimenter's clock; independent of coordinate system (Kenyon 1991).

[5] Mach's principle is essentially that inertial forces are a result of the distant stars.

$2GM/rc^2$ comes from the fact that the proper time is squared, measured from the coordinate time squared, times the difference of gravitational potential, being 2Φ.

Later in 1915, Karl Schwarzschild derived an exact solution to Einstein's GR field equations, published in 1916 (Schwarzschild, 1999). See Appendix 1.1.3 for details. The Schwarzschild solution is used in *most* tests of GR unless otherwise stated. He also derived the 'Schwarzschild radius,' $R_s$, which is the radius of a sufficiently massive object where all particles, including photons, which will inevitably fall into a massive central object (Schwarzschild, 1999). This is shown in Equation 7, while the origin of the $2GM/c^2$ was revealed in Equation 6.

$$R_s = \frac{2GM}{c^2} \qquad \text{Eq. 7}$$

Now a weak−field[6] regime and a strong−field regime will be defined. The gravitational field in the solar system is weak. Hence it is in the weak−field regime, or post−Newtonian regime. Assuming the field is quasi-stationary, it can be described by a symmetric tensor in flat spacetime, using the Minkowski metric[7] (Schutz, 2011). However, it is convenient to simply define the *strength* of a gravitational field using Schwarzschild spacetime. The distance from the gravitating object *r* and its mass *M* are arranged as follows in Equation 8.

$$\epsilon \equiv \left[\frac{v}{c}\right]^2 \equiv \frac{GM}{rc^2} \qquad \text{Eq. 8}$$

The *strength* of the gravitational field is $\epsilon$, being the ratio of the gravitational potential energy of a test particle and its rest mass energy. When the gravitational field is weak; then, $\epsilon \ll 1$, typically $\epsilon < 10^{-5}$, as is the case within the solar system. The strong−field regime is generally near very massive objects. Moreover, the strongest fields are when $\epsilon \to 1$. This occurs as the event horizon of a non-rotating black hole is approached or the expanding observable universe. At the surface of neutron stars $\epsilon \sim 0.2$, again in the strong−field regime (Will, 2014).

Many GR tests report using the parameterized post-Newtonian [PPN][8] framework[9], and so this will be introduced briefly. PPN formalisation provides a simple method of describing deviations from GR. Two parameters are specifically used to test GR. The first one is $\gamma$, being the amount of curvature produced, per unit mass. Secondly, Dicke (1961) stated that ratio of inertial mass to gravitational mass, for astronomical objects may differ from unity. When an object varied its position in the gravitational field of another object; then, this is the proposed nonlinear aspect of gravity and is denoted by, $\beta$. This is also referred to as nonlinearity of gravity (Williams et al., 1976; Nordtvedt Jr, 1982). Both these parameters in GR should be unity. There are other PPN parameters, producing a more complex relation, but in GR these are zero and are generally used to test other models (Merkowitz, 2010; Dicke 1961). See Appendix 1.1.4 for more details of the PPN values.

[6] Often referred to as 'linearized gravity.'

[7] The Minkowski metric is a 4D flat spacetime metric, which is fundamental to the special theory of relativity.

[8] PPN: parametrized post-Newtonian notation expresses general relativity equations in terms of the lowest-order deviations from Newton's law of universal gravitation, often used in the case of weak-fields.

[9] There are other frameworks, such as the Robertson-Mansouri-Sexl framework, but will only consider the PPN.

Additionally, other parameters that are used in tests of GR are the implementation of Post-Keplerian [PK] parameters. See Appendix 1.1.5 for more details.

This chapter will now explore the three classical tests of GR.

**1.2 Perihelion Advance of Mercury**

The first general relativistic effect, to be observed, was the advance of the perihelion in Mercury's orbit; it was viewed by Le Verrier in 1859 (Le Verrier, 1859). The advance of the perihelion sometimes referred to as apsidal precession or orbital precession, basically means the orbit is precessing as shown in Figure 2. It would be extremely imprecise to measure the precession of one orbit, but measuring the accumulated perihelion advance over 100 years, makes the effect measurable. It is measured in arcseconds per century, and ascribed the following notation, $\Delta\phi$ century$^{-1}$.

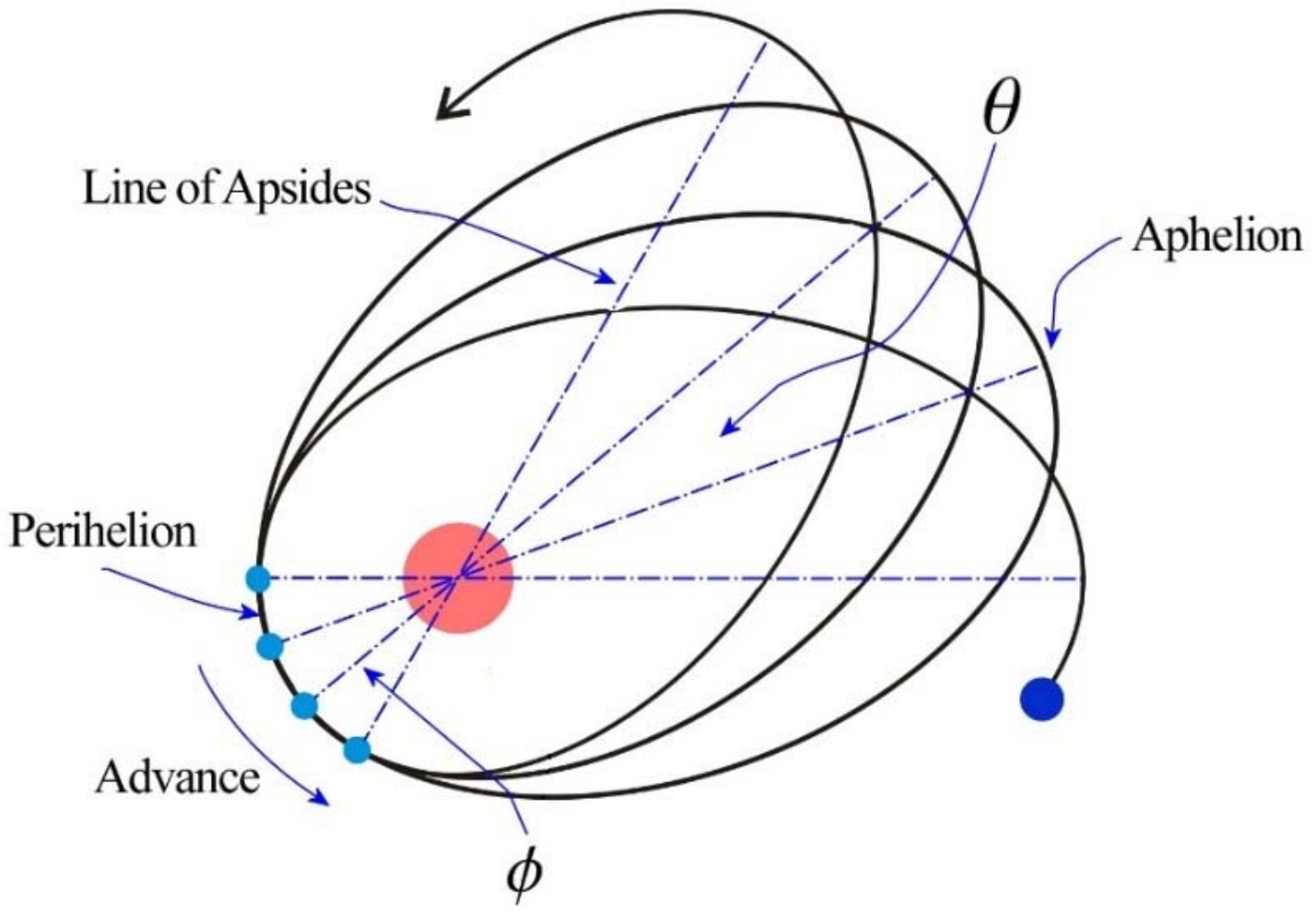

*Figure 2: Precession of the perihelion, after one orbit makes the angle $\phi$. The effect is greatly exaggerated. Illustration by E. Asmodelle.*

The Newtonian treatment for value for apsidal precession is concerned with forces, as is the law of universal gravitation. According to Newton's 'Principia,' Proposition 45 of Book I, the centripetal force could be represented as shown in Equation 9, for orbits almost circular; namely, the eccentricity $e \ll 1$. The value of the constant $\mu$ is often taken to be ~1. The index for the centripetal force $n$ is a function of varying $e$. When there was no precession $n = 1$, then the value is *quiescent*, as the form resorts to the inverse square law. Hence, for apsidal precession $n \neq 1$ (Valluri et al., 1997). If $e \sim 0$, then apsidal precession would *not* occur.

$$\mu r^{n-3} \qquad \text{Eq. 9}$$

The angle *between* lines of apsides is $\theta$ and is provided in Equation 10; also note: $\theta \equiv \phi$.

$$\theta = \pi/\sqrt{n} = 180°/\sqrt{n} \qquad \text{Eq. 10}$$

In Newton's, 'System of the World,' Book 3, he states that precession can be accounted for by perturbation (Harper, 2007). The work *does* predict apsidal precession. However, following users of this method found that it failed to account *fully* for the phenomena (Valluri et al., 2005).

Le Verrier[10] used perturbation theory[11] and planetary transit data, spanning 50 years. He calculated the interaction of Mercury with the other planets and found the residual of 39" century$^{-1}$ was not accounted for by Newtonian perturbations (Treschman, 2004). In 1882, Newcomb determined the rate of precession to be 531" century$^{-1}$, leaving a residual of 43" century$^{-1}$ less than the observed rate, using similar methods (Harper, 2007). Theories emerged to resolve the residual, using Newtonian equations, but they were ad hoc and failed to predict much else. People speculated that the excess effect, or residual, was due to an unseen planet, they named *Vulcan*[12] or some other collection of objects.

In 1915 Einstein derived the correct value of the precession of the perihelion of Mercury (Einstein, 1915b). The fact that other solar system planets have a significant gravitational effect on Mercury, albeit tiny, meant Mercury did not move in precisely a $1/r$ orbit, being the Newtonian potential (Hartle, 2003). The GR equation of motion, or *orbital* equation, is provided in Equation 11. See Appendix 1.2.1 for the Newtonian equation of motion.

$$\frac{d^2u}{d\phi^2} + u = \frac{GM}{h^2} + \frac{3GM}{c^2}u^2 \qquad \text{Eq. 11}$$

---

[10] Le Verrier had previously discovered Neptune from its gravitational effect on the orbit of Uranus.

[11] Perturbation theory is the study of dynamical systems that have small perturbations of simple or linear systems.

[12] 'Vulcan' was a hypothetical small planet that was believed to exist in an orbit between Mercury and the Sun.

Using the notation of Hobson et al., (2006): the angular momentum, per unit mass, of the orbiting particle is $h$, $u \approx GM/h^2 \{1 + e \cos [\phi (1 - \Delta u)]\}$, where $\Delta u$ is the perturbation: $\Delta u = A [1 + e^2 (\frac{1}{2} - \frac{1}{6} \cos 2\phi) + e\phi \sin\phi]$ and $A = 3(GM)^3/(c^2 h^4)$, which is tiny. The residual $\Delta\phi$, can then be derived as shown in Equation 12, while the semi-major axis $a = 5.8 \times 10^{10}$ m, solar mass is $M_\odot = 2 \times 10^{30}$ kg, and $e = 0.2$.

$$\Delta\phi = \frac{6\pi GM}{a(1 - e^2)c^2}$$
Eq. 12

Einstein's GR predicted value for $\Delta\phi$, was 43″ century$^{-1}$, while astronomers at that time determined $\Delta\phi$ to be, 45″ ± 5″ (Einstein, 1915b). Recently the observational value of $\Delta\phi$ has been constrained to 42.”98 ± 04 century$^{-1}$ (Lo et al., 2013). See Appendix 1.2.2, for a complete equation for $\Delta\phi$.

**1.3 Deflection of light**

Einstein's work of 1907 also revealed that light was bent as a result of spacetime curvature (Einstein, 1907). In the general relativistic case, this is only true of a stationary observer who sees the light pathway relative to a gravitating body. Einstein understood, using the EEP, that mass, or indeed energy from $E = mc^2$, would follow geodesic pathways in curved spacetime as depicted in Figure 3, relative to an observer at rest with the gravitating body.

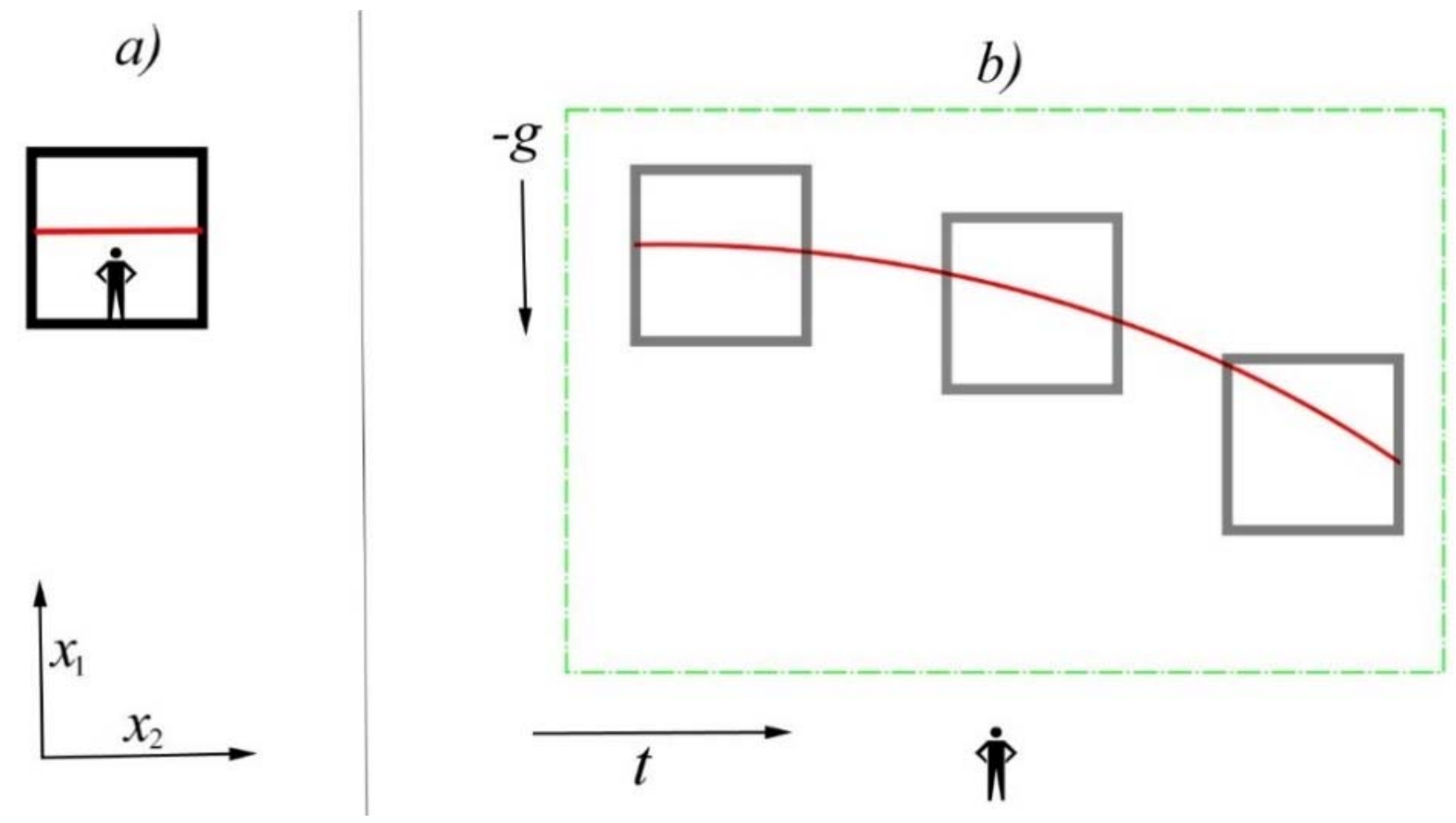

*Figure 3: Part a) shows the reference frame, that is at rest or in free fall, where light is being emitted. Part b) shows the same event, relative to an observer who sees the emitted light move as a result of gravity, in curved spacetime. Illustration E. Asmodelle.*

This concept embodies the essence of EEP, showing that gravity and acceleration are indistinguishable from each other, within a small region. As the speed of light is constant, Einstein also showed that time dilation occurs in the reference frame where the light was produced. In part b), time has to run slower, or become dilated, to match the coordinate time of the reference frame in a). Meaning, light had to travel further without going greater than *c*.

Assuming that light from a distant star is passing by the Sun or from any gravitating body, at its closest approach, Equation 13 represents the GR equation of motion of deflection.

$$\frac{d^2u}{d\delta^2} + u = \frac{3GM}{R_0^2} cos^2\delta$$

Eq. 13

The radius of the Sun is, $R_0$, and $u \equiv 1/r$, being due solely to the Sun and is the Newtonian potential. The angle of deflection is $\delta$ (Narlikar, 2010). Note the similarity with Equation 11.

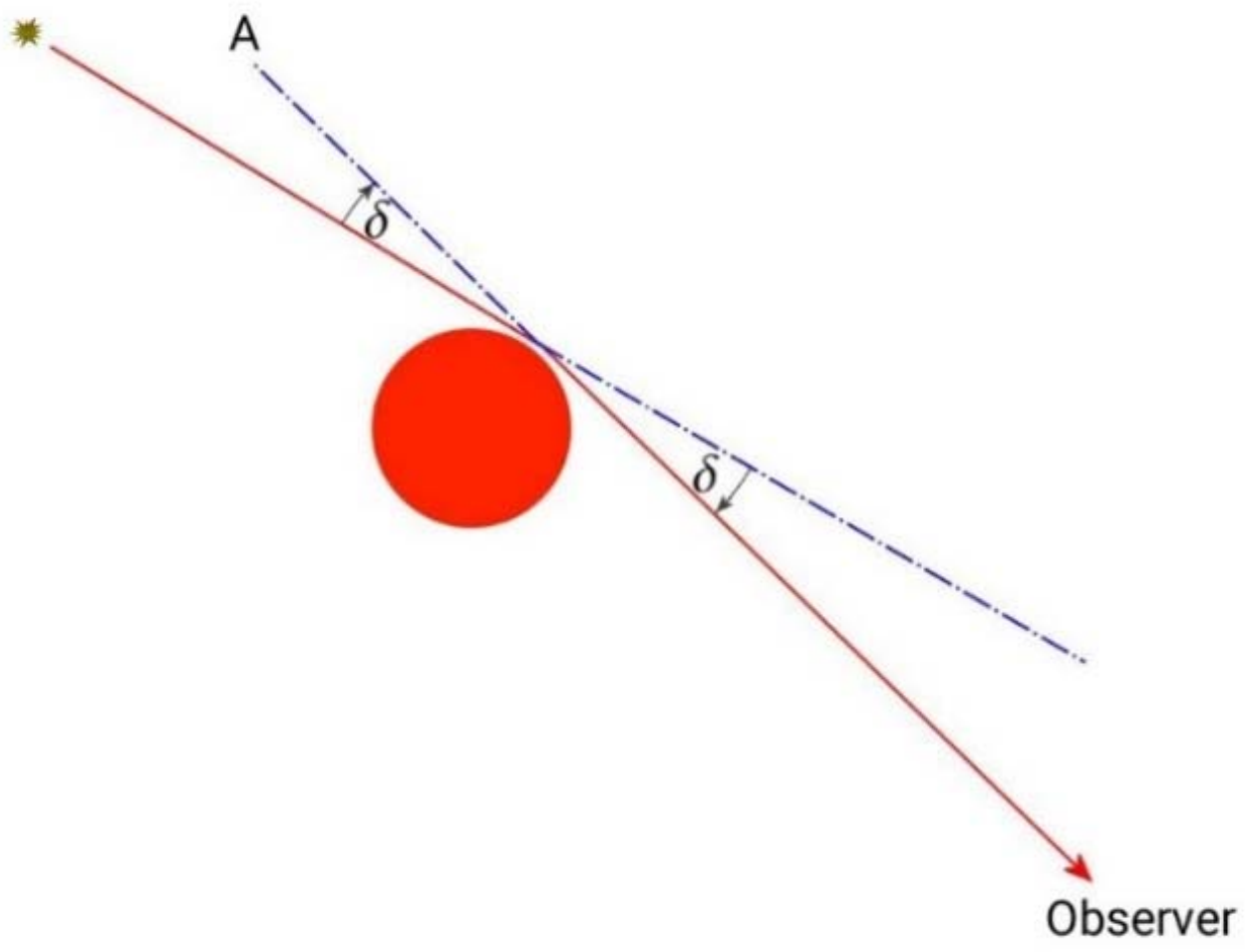

*Figure 4: The bending of light by a massive object is, in this case, the Sun. Point A is the apparent location of the source as seen by the observer. Illustration E. Asmodelle.*

After some manipulation and a few assumptions, the net deflection of light according to GR $\delta_{GR}$ reduces to Equation 14, as depicted in Figure 4.

$$\delta_{GR} = \frac{4GM}{R_0c^2}$$

Eq. 14

In contrast, the Newtonian result for the deflection of light $\delta_{NG}$ is shown in Equation 15 (Soares, 2009). The results were such that, $\delta_{NG}$ resulted in ~0".87 and $\delta_{GR}$ resulted in ~1".75.

$$\delta_{NG} = \frac{2GM}{R_0c^2} = \frac{1}{2}\delta_{GR}$$

Eq. 15

Einstein suggested, in his 1911 paper, that *this* phenomenon may be testable[13] during a solar eclipse (Einstein, 1911b). Then in 1917, Sir Frank Dyson the Astronomer Royal, and Arthur Eddington planned to photograph a total solar eclipse on May 29, 1919, as the sun was crossing the Hyades cluster. Eddington measured the true positions of Hyades between January and February of 1919. Then in May 1919, he travelled to the island of Príncipe, off the west coast of Africa, situated in the Gulf of Guinea[14]. There, during that total solar eclipse, he measured the apparent change in position, or displacement, of the stars in the Hyades cluster as a result of the Sun's gravitational lens. The stars nearest the Sun's rim appeared to move outwards radially from the Sun as depicted in Figure 5.

[13] There had been four unsuccessful attempts to photograph the effect during a solar eclipse.

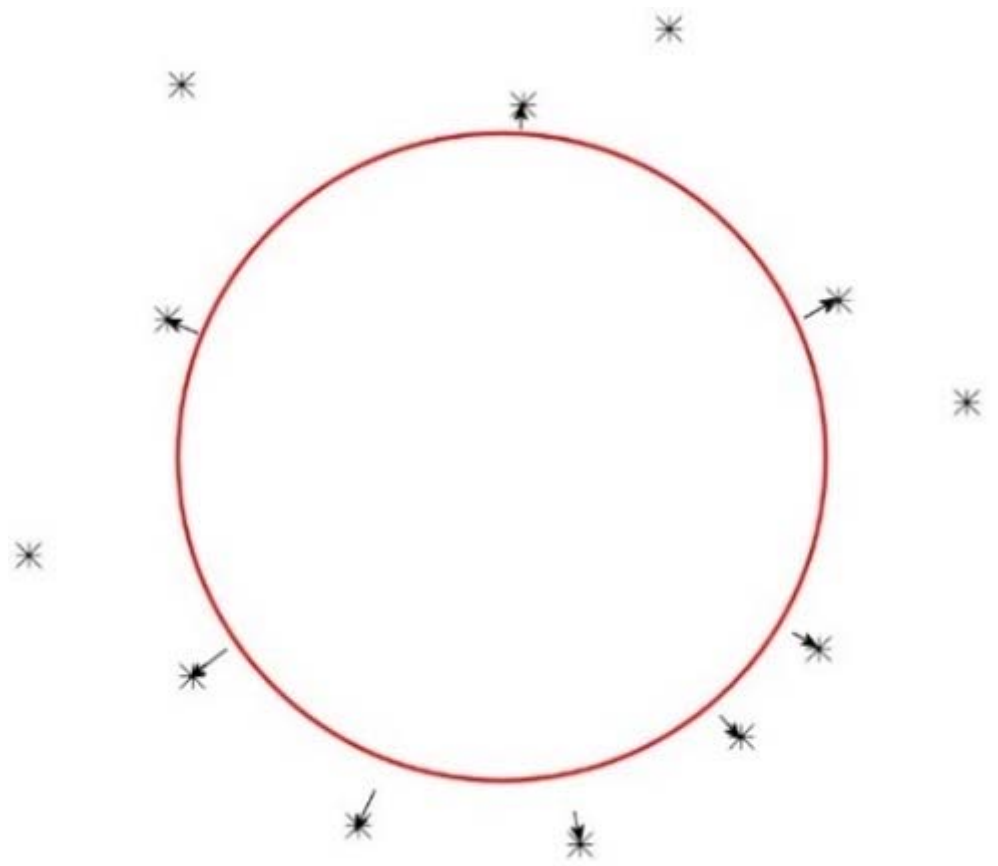

*Figure 5: When photons from stars are passing close by the Sun, as in a solar eclipse, their apparent position seems to move radially outward. Illustration E. Asmodelle.*

On November 6th, 1919 Eddington presented his findings. The results of two different photographic plates of the displacement were stated: 1".98 $\pm$ 0".12 and 1".61 $\pm$ 0".30. Eddington said that the predictions of GR were compatible with the results of their observed deflection, whereas the Newtonian was not (Dyson et al., 1920). The theoretical prediction of GR was 1".75. However, there are other factors to consider[15].

In succeeding years, more measurements were made, but there was no significant improvement until the advent of very-long-baseline interferometry [VLBI] in 1972. VLBI is the simultaneous use of various radio telescopes that are situated at distant locations from one another which act as an interferometer. This technique creates a virtual radio telescope with a far larger baseline. VLBI uses a geometric method to measure the arrival time of radio waves from an astronomical source between different telescopes $\tau$, and their angular separation $\beta$ and a geometric measure of the projected baseline U. The VLBI principle is shown in Figure 6 with two radio telescopes. The signal delay L is a trigonometric relationship between those variables and the baseline B.

[14] Eddington, as a backup, also sent a group of astronomers to Sobral, Brazil to photograph the eclipse.
[15] Refraction by the Sun's outer atmosphere, optical distortion and other factors.

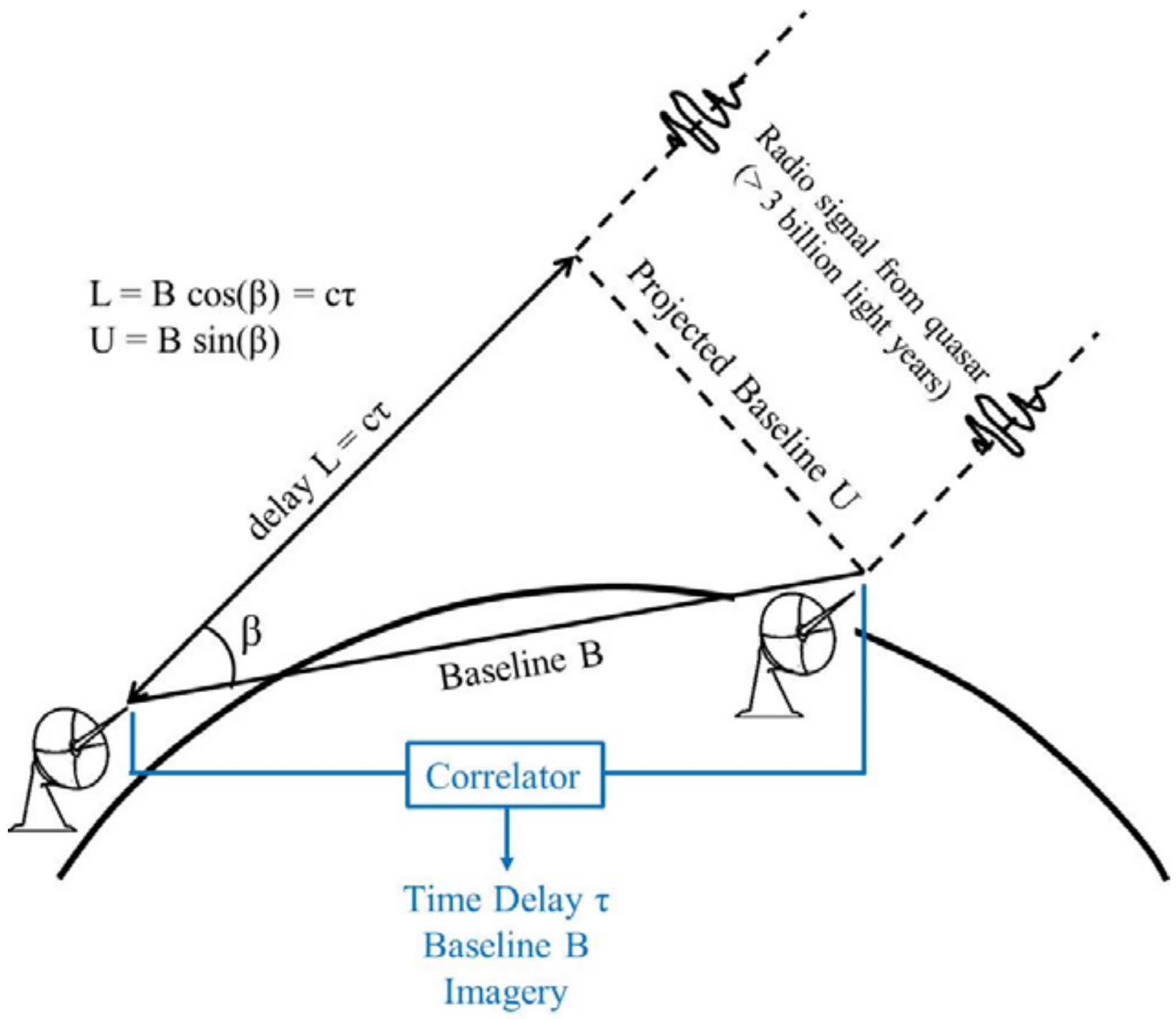

*Figure 6: VLBI detecting radio signals from a quasar.*
*Courtesy: Computational physics Inc.,* http://www.cpi.com/projects/vlbi.html

Shapiro et al., (2004) reports on deflection by the Sun of radio waves emitted from extragalactic radio sources, between 1979 – 1999. Using VLBI for radio band detection provides a greater level of accuracy over the visible band, for the visible band is degraded by atmospheric absorption and scintillation. The PPN [defined p. 8] framework results were, $\gamma = 0.9998 \pm 0.0004$, compared to the GR value of, $\gamma = 1$.

These tests of the bending of light, or gravitational lensing, are observational evidence that photons follow geodesics in spacetime, being a fundamental precept in GR. This was an early monumental verification for GR.

### 1.4 Gravitational Redshift

In 1925 Adams measured the gravitational redshift $z_g$, of Sirius B[16], which was non-Doppler and measured to be $z_g = +23$ km s$^{-1}$, yet uncertainties were not stated (Adams, 1925). This was the first observational verification of gravitational redshift [GvR], yet Adams's measurement was only one quarter the actual value. More recently, the Hubble Space Telescope [HST], measured the value of GvR of Sirius B to be, $z_g = +80.42 \pm 4.83$ km s$^{-1}$ (Holberg, 2010).

[16] Sirius B is a white dwarf is at ~ 2.64 pc, hence because of its proximity it is free of cosmological redshift.

In 1959 Pound and Rebka, devised a precision method to experimentally measure the GvR, using the Mössbauer effect (Pound & Rebka, 1959). When a photon is emitted or absorbed, an exchange of momentum with the nuclei occurs, recoiling the nucleus which reduces the emitted photon's net energy, hence its frequency. This is a result of the conservation of momentum. Rudolf Mössbauer in 1958, determined that an almost recoil-free emission and absorption of gamma radiation can occur by embedding atomic nuclei in a solid lattice crystal so that it was resistant to recoil. This is the Mössbauer effect and would permit researchers to detect the GvR effect without contamination by the recoil of the nucleus during absorption or emission, and reduced energy of emitted photons (Eyges, 1965).

In the following year, 1960, Pound and Rebka arranged their experiment in a high tower at Jefferson laboratory, on the Harvard campus. The physical setup is depicted in Figure 7. The height of the experiment is $h$, being 22.6 m. The fractional width and intensity of Fe spectral lines seemed sufficient to measure the GvR effect in the laboratory and was therefore selected. The source, excited Fe ion which released gamma photons was embedded in a lattice.

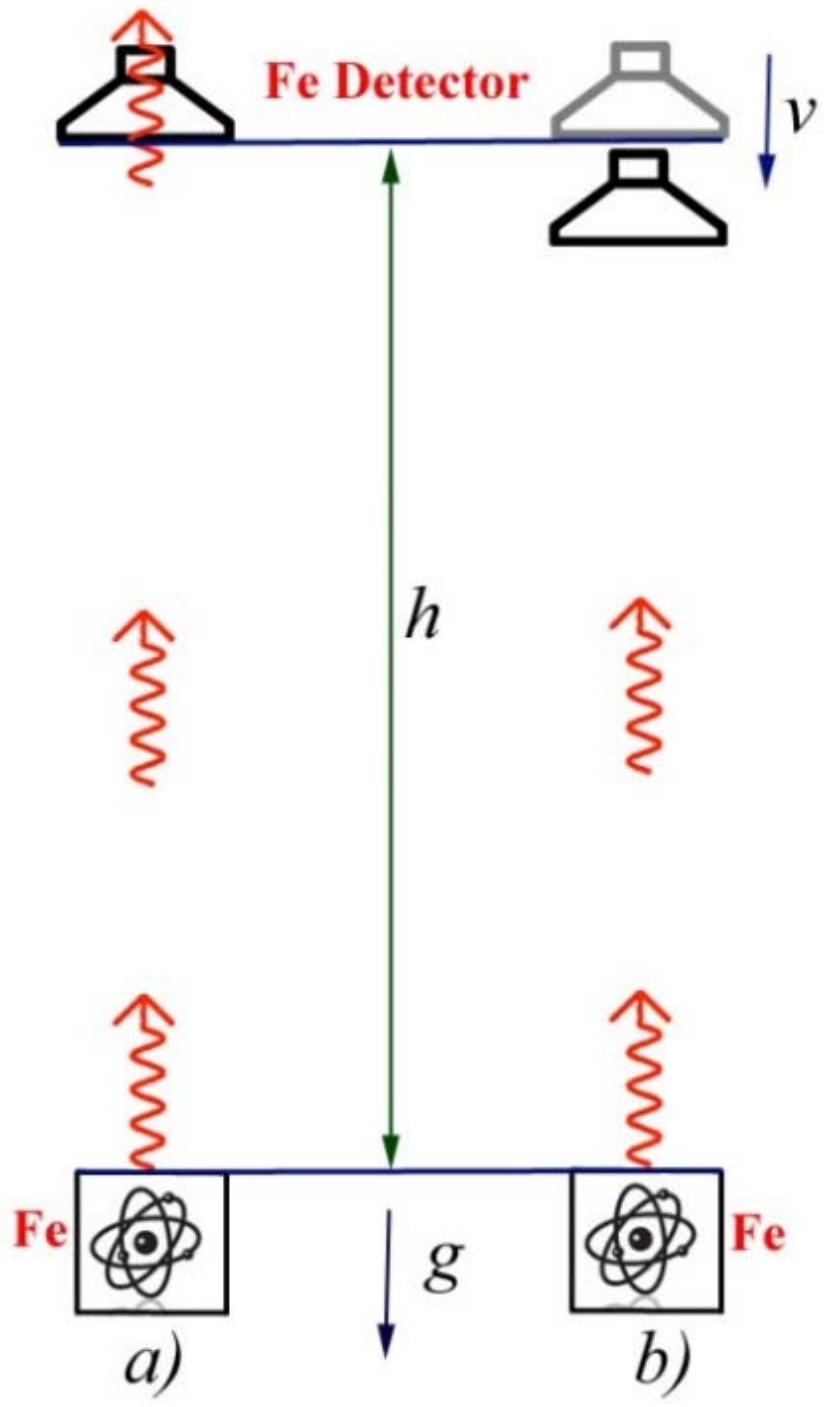

*Figure 7: This is an illustration of the Pound and Rebka experiment when in actual fact the experiment was also conducted in an inverted configuration. In part a) the detector is stationary and the photons pass right through. In part b) the detector is moving with velocity* v *and the photons are absorbed. Illustration E. Asmodelle.*

As the photon climbed upward against the gravitational field of the Earth $g$, the photon became gravitationally redshifted. When the photon arrives at the Fe ion detector, it has a lower frequency as seen in part *a)*, and was not absorbed. In part *b)*, the Fe ion detector is moved towards the emitter; when the Doppler blueshift of the moving emitter is the same in magnitude, but opposite in sign, as the GvR effect; the photon has the right energy and is absorbed.

To an observer in *b)* at rest, the falling detector has increased its speed, as in Equation 16.

$$g\Delta t = gh/c$$ Eq. 16

Hence, the velocity of the detector is, v = $g\Delta t$. When travelling towards the Earth the effect is a blueshifted Doppler shift which is depicted by Equation 17.

$$\Delta\lambda = \lambda(\mathrm{v}/c) = \lambda(gh/c^2)$$ Eq. 17

Therefore, the gravitational redshift $z_g$, can then be represented by Equation 18, where the gravitational potential of the Earth is $\Phi_E$.

$$z_g = \frac{gh}{c^2} = \frac{\Delta\Phi_E}{c^2}$$ Eq. 18

This experiment was repeated many times and achieved the first physical verification of GvR, measuring a net fractional shift of ~ $-(5.13 \pm 0.51) \times 10^{-15}$. In contrast to the GR estimated value of, ~ $-(4.92 \pm 0.51) \times 10^{-15}$ (Pound & Rebka, 1960). The precision of this measurement was independent of height, while later versions of the experiment constrained the result to within ~1% of the GR prediction.

This result was a huge scientific achievement for experimental testing of GR, not only because it used the Mössbauer effect and produced a high-resolution result, but also because of the unique and ingenious experimental arrangement. The accuracy of these tests is comparable to the precision of modern ones.

Many experimental tests of GvR followed, while some used hydrogen masers in space. A Scout rocket was launched in 1976, which carried the Vessot–Levine Experiment as part of Gravity Probe A [GP-A]. GP-A was a NASA a Space-Based geodetic satellite to test GR. This experiment constrained the relative discrepancy with GR to ~$70 \times 10^{-4}$ level (Vessot et al., 1980; Will, 1993). The Russian satellite RadioAstron reduced the relative discrepancy to ~$10^{-6}$ (Biriukov et al., 2014). These latter results have constrained GvR to high levels of precision.

Lastly, the tests of GvR are classified as the third classical test of GR.

# Chapter 2:

## Modern tests

### 2.1 Shapiro delay

The light travel time delay, or Shapiro delay, is sometimes called the *fourth* classical test of GR and was first introduced by Shapiro in 1964. This delay in the arrival time of light passing nearby a massive object is the casual result of general relativistic time dilation, due to a significant gravitational potential. Shapiro realised, the speed of light is dependent on the gravitational potential along its travel path, according to GR, and could be tested with a close conjunction of the Sun (Shapiro, 1964). These tests were measured in the weak−field regime.

The physical definition of $r_p$, $r_e$ and $r_0$, during a test scenario are represented by Figure 8.

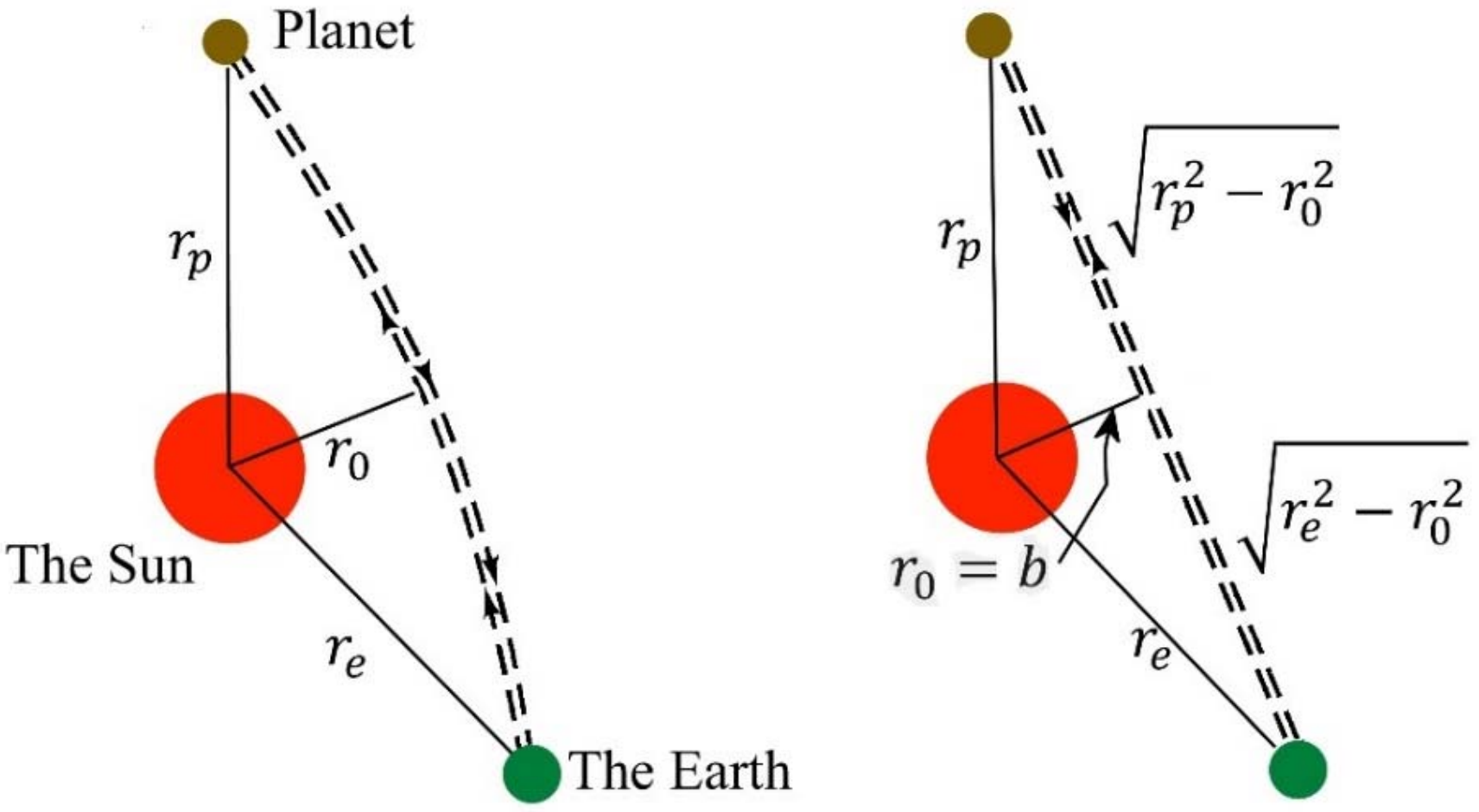

*Figure 8 The radar reflection of photons from the Earth to a planet and back. The left image is the actual path, exaggerated. The right image is the Euclidean form. Illustration E. Asmodelle.*

To define Shapiro delay, assume the Earth and the planet are stationary, while the total time for the round trip of the radar signal is $\Delta t$, in coordinate time. The value of $t$ must be represented in terms of $r$ along the entire pathway, while $r_0$ is the closest approach to the Sun or its radius. The travel time from Earth to the reflecting planet and back to Earth is given in Equation 19, where $k$ is the Schwarzschild metric parameter: $k = GM_\odot / c^2$. Also $r_e$ is the orbital radius of the Earth and $r_p$ is the orbital radius of the reflecting planet[17] (Shapiro, 1964; Lambourne, 2010).

$$\Delta t \approx \frac{2}{c}\left[\left(r_e^2 + r_0^2\right)^{1/2} + \left(r_p^2 + r_0^2\right)^{1/2}\right] + \frac{4k}{c}\left\{\ln\left(4\frac{r_e r_p}{r_0^2}\right) + 1\right\} \qquad \text{Eq. 19}$$

[17] Of course, corrections for the movement of the planets and the refraction by the Solar corona, must be taken into account.

This leads to a viable solution for the excess time delay, or Shapiro delay $\Delta t_{excess}$, provided in Equation 20. An alternative notational derivation is presented in Appendix 2.1.1.

$$\Delta t_{excess} \approx \frac{4GM}{c^3}\left[\ln\left(\frac{r_p r_e}{r_0}\right)+1\right] \quad \text{Eq. 20}$$

In 1965, Shapiro et al., using the Lincoln Laboratory Haystack radar system, began a program to measure the delay in reflected radar signals off solar system planets. The radar system had ~200 $\mu$sec accuracy. In 1966, they measured the Shapiro delay of signals bounced off Venus, and also from Mercury in 1967, both during a super-conjunction when the Shapiro delay signal is the *strongest* as the gravitational potential is *highest*. See Figure 9, for the Mercury data. The time delay corresponded with GR predicted value to within, ~20% (Shapiro et al., 1968).

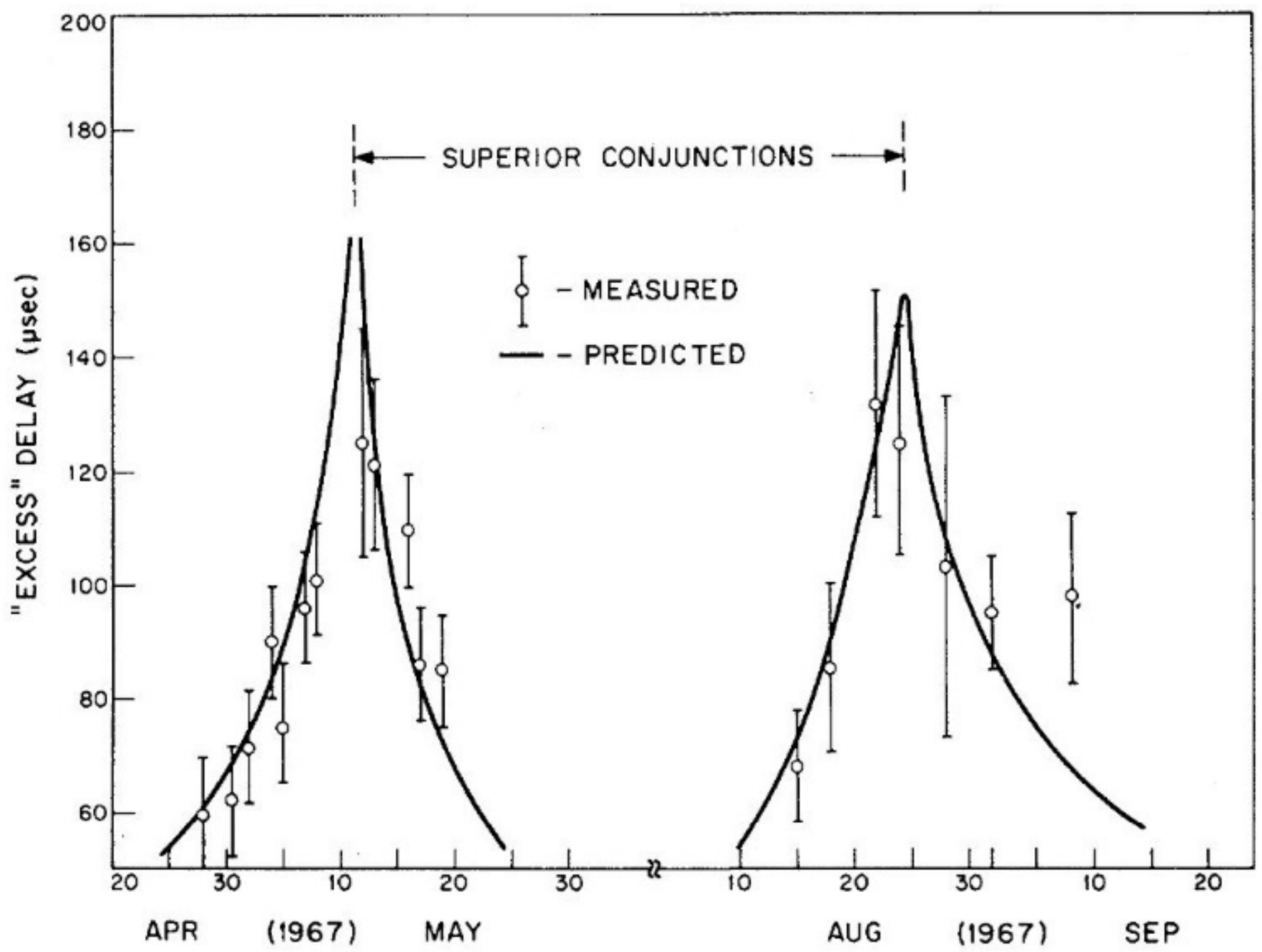

*Figure 9: shows measured and predicted Shapiro delays, in μsec, based on GR for Earth-Mercury. (Shapiro et al., 1968).*

Subsequent measurements in 1971 further reduced the deviation from GR to ~ 5%. Then in 1979, transponders were placed on the Viking spacecraft, whereby radar signals were bounced off Mars from the Viking lander. This enabled Shapiro and Reasenberg to improve the accuracy of the Shapiro delay results, to better than ~0.1% with GR (Rindler, 2006).

In 2003, the Cassini spacecraft flew by Saturn and measured the Shapiro time delay of bounced signals. The deviation of the PPN parameter $\gamma$, being the amount of curvature produced per unit mass, was 1.000021 ± 0.000023, against the GR value of $\gamma = 1$ (Ni, 2005). This result confirmed the GR effect of Shapiro delay to approximately 20 parts per million (Cooperstock, 2009).

Fundamentally, Shapiro time delay was not originally thought of by Einstein and his contemporaries, and so its later introduction and experimental verification elucidated GR.

### 2.2 Gravitational time dilation

Gravitational time dilation [GvT] is distinctly different from time dilation in special relativity [SR][18] and also cosmological time dilation[19]. Equation 3 only provides the first order of GvT, whereas the exact form is shown in Equation 21. This provides the small element of proper time $d\tau$, being dilated from the coordinate time $dt$, near a massive object[20] (Hobson et al., 2006).

$$d\tau = \left(1 + \frac{2\Phi}{c^2}\right)^{1/2} dt \qquad \text{Eq. 21}$$

Expanding this relation, with the potential $\Phi$ depicted in Equation 1; namely $\Phi = -GM/r$ provides Equation 22. This asserts that when $\Phi \rightarrow 0$, then $d\tau \rightarrow dt$. Hence, when there is no gravitational potential[21], there is no time dilation.

$$d\tau = \left(1 - \frac{2GM}{rc^2}\right)^{1/2} dt \qquad \text{Eq. 22}$$

The first direct experimental confirmation of GvT was performed in 1971 by Hafele and Keating, who flew four Cs-clocks[22] around the world, both eastward and westward. Clocks in the plane lost $-59 \pm 10$ ns travelling eastward, and gained $+273 \pm 7$ ns travelling westward. The respective GR predictions were $-40 \pm 23$ ns; $+275 \pm 21$ ns (Hafele & Keating, 1972a). This too, being in Earth's orbit, was measured in the weak−field regime. The resulting total time dilation was in agreement with SR and GR, to within ~10%. In the Hafele and Keating experiment, the total time dilation $\tau$, for one trip, is represented by Equation 23, where $\tau_0$ is the initial time, and the altitude is $h$.

$$\tau - \tau_0 = \left[gh/c^2 - \left(2R\Omega \mathrm{v} + \mathrm{v}^2\right)/2c^2\right]\tau_0 \qquad \text{Eq. 23}$$

A stationary equatorial clock on Earth, relative to non-rotating space, has a 'clock rate' of $R\Omega$, while v is the velocity, $gh$ is the gravitational potential, and $gh/c^2$ is the GvR term (Hafele &

[18] In special relativity, observers in relative motion experience *time dilation*; a consequence of non-simultaneity.

[19] In an expanding universe, *cosmological time dilation* occurs when the emitted light from a distant source at redshift $z$, is dilated, to a local observer, by a factor $\sim 1 + z$.

[20] This representation is different to Einstein's original Equation 3, being a formula for interval time, and not proper time as in Equation 21 and 22.

[21] This refers to the gravitational potential in excess of the observer's gravitational potential.

[22] A Cs-clock is a caesium standard or caesium atomic clock.

Keating, 1972b). The results are shown in Figure 10. Many subsequent tests of this kind, have been carried out, further reducing the uncertainty to ~ 1.6% (Alley, 1979).

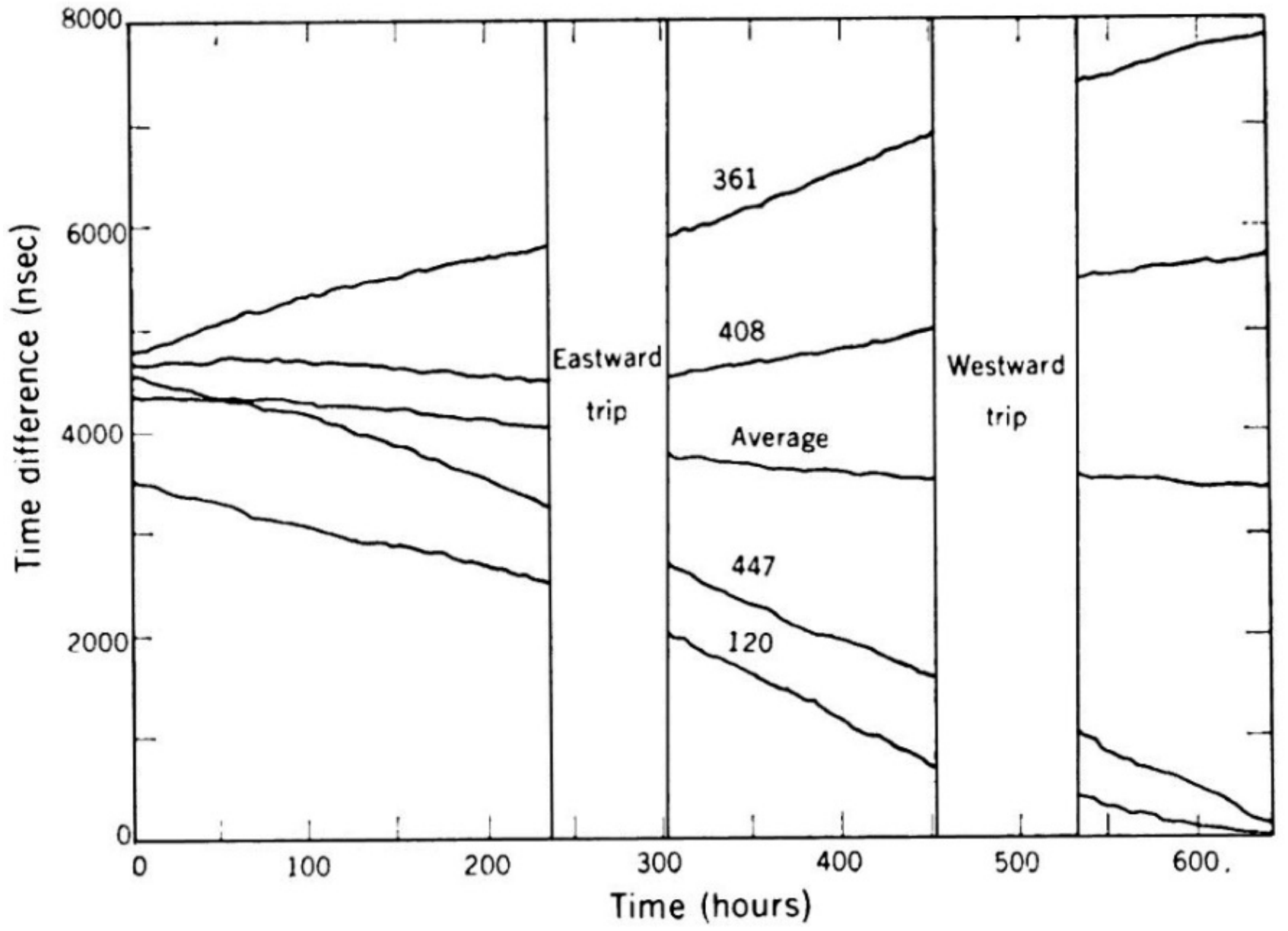

*Figure 10: Hafele and Keating experiment, showing the time differences between each clock at hourly intervals. Courtesy: (Hafele & Keating, 1972a).*

A similar test was performed, with the first FRG Spacelab Mission Dl, on the Space Shuttle Challenger in 1985. The on−board NAVEX Cs-clock, against the ground reference clock, determined GvT to ~ $5 \times 10^{-4}$, with a l$\sigma$ uncertainty[23] (COSPAR, 1988).

The Global Positioning System [GPS][24] basically consists of 24 satellites with on−board atomic clocks, is a daily validation of SR and GR. GPS calculations utilize the Kerr metric[25], see Appendix 2.2.1. The GPS satellite system runs on Universal Coordinated Time [UTC], and is managed by the U.S. Naval Observatory [USNO]. The determination of transfer time and position is represented by Equation 24. For example, using $n$ synchronized atomic clocks, transmitting pulses from their positions at distances $\mathbf{r}_j$, and at times $t_j$, where $j = 1, 2,..n$. These signals are received, at a given satellite, at a distance from the Earth by the position vector $\mathbf{r}$, at time $t$ (Ashby, 2003), due to the constancy of the speed of light, the following relation holds.

$$c^2(t - t_j)^2 = |\mathbf{r} - \mathbf{r}_j|^2 \qquad \text{Eq. 24}$$

See Figure 11 for a visual representation of Equation 24, where $n = 4$.

[23] 1$\sigma$ (sigma), corresponds to one standard deviation, which is a confidence that the result is real to ~ 66%.

[24] These relativistic effects also apply to the Global Navigation Satellites Systems [GNSS] and other systems.

[25] The Kerr metric, considers the central object to be rotating, opposed to stationary in the Schwarzschild metric.

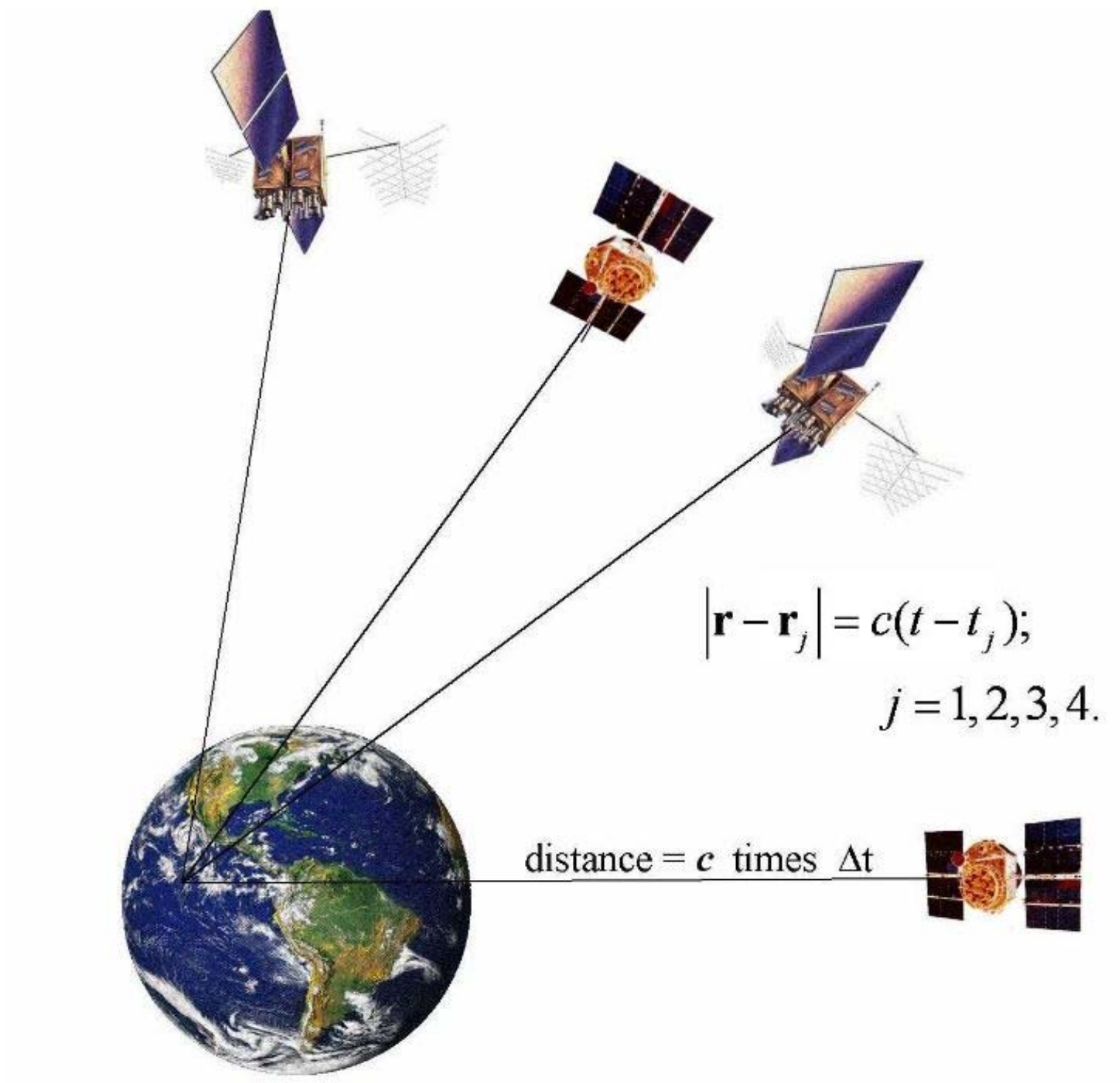

*Figure 11: The positions of receivers are shown. Transmitted at $t_j$, and at positions $r_j$. The values of GvT is found by solving four simultaneous equations. Courtesy: (Ashby, 2002).*

The GPS has various relativistic effects; namely, time dilation due to SR, GvT, and GvR, as well as frame−dragging, and the Sagnac effect[26]. Other errors to be accounted for are provided in Appendix 2.2.2. However, our interest here is purely the GvT value due to the GR, according to Zhao et al., this value is, ~ $4.5 \times 10^{-10}$, or 4.5 in $10^{10}$ parts (Zhao et al., 2011), being in the weak−field regime.

## 2.3 Frame−dragging & Geodetic effect

Other phenomena that are predicted by GR are the 'frame–dragging' of spacetime or Lense–Thirring precession, and the 'geodetic effect.' These are discussed respectively. These tests utilize the Kerr metric, as does the GPS.

### *Lense–Thirring precession*

Lense–Thirring precession [LTP][27] is distinctly different from de Sitter precession[28], having been tested successfully marks an important step in GR's confirmation. The dragging of inertial frames by a massive body is a direct consequence of GR. Although, it was Lense and Thirring in 1918, who derived the equations of frame–dragging, using the Kerr metric (Lense & Thirring, 1918).

[26] The Sagnac effect is the phenomenon of interference when a ring interferometer undergoes rotation.

[27] The Lense–Thirring effect is a global effect of the entire orbit, while the Schiff effect follows the Fermi-propagation of spin of a gyroscope (Blanchet et al. 2011).

[28] The de Sitter precession is due to the presence of a central non-rotating mass.

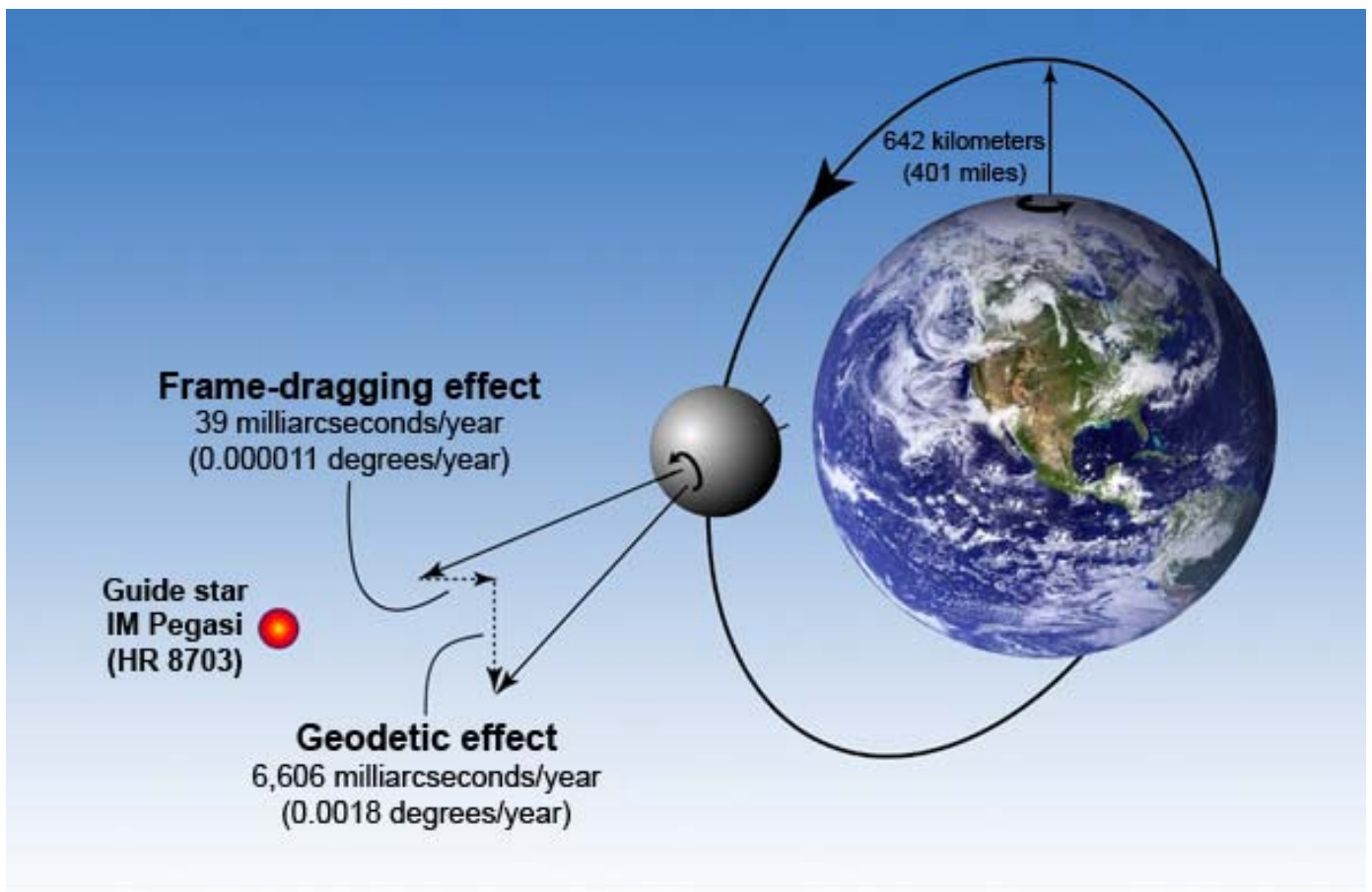

| | Measured | Predicted |
|---|---|---|
| Geodetic precession (mas) | 6602 ± 18 | 6606 |
| Frame-dragging (mas) | 37.2 ± 7.2 | 39.2 |

*Figure 12: shows two effects: 1) Frame-dragging, and 2) the Geodetic effect. Courtesy (Will, 2011) https://physics.aps.org/articles/v4/43*

LTP can be seen in a satellite in a polar orbit around the Earth, as the Earth undergoes precession[29] about its axis, through the poles. LTP predicts that a satellite's plane of orbit will precess slowly in the direction of the Earth's rotation (Lambourne, 2010). The effect is depicted in Figure 12.

LTP is represented by Equation 25. The value $I$ is the moment of inertia and $\omega$ is the angular velocity, both of the central mass: the Earth. Also, $a$ and $e$ are the semi-major axis and eccentricity of the orbiting body, respectively. The unit vector **n** is parallel to the central body's spin axis. Hence, the resultant is the advance rate of precession Ω (van Patten & Everitt, 1976).

$$\boldsymbol{\Omega} = \frac{2GI\omega}{c^2 a^3 (1-e^2)^{3/2}} \mathbf{n}^2 \qquad \text{Eq. 25}$$

Ciufolini and Pavlis in 2004, reported measuring LTP on the LAGEOS[30], LAGEOS2, and GRACE[31] to 99 ± 5% of that predicted by GR, with a 10% overall uncertainty (Ciufolini &

[29] Precession is a product of tilting, the equatorial bulge, influence of the Sun, Moon and to a lesser extent the planets, just as a spinning top precesses due to the effects of gravity (Dehant & Mathews 2015).

[30] LAser GEOdynamics Satellite, or LAGEOS 1 and 2 were launched in 1976 by NASA.

[31] GRACE was the second mission in NASA's Earth System Science Pathfinder (ESSP) Program in May 1997.

Pavlis, 2004). This uncertainty was later reduced to 0.994 ± 0.002 to the value of GR (Ciufolini et al., 2016).

***Geodetic precession***

The next phenomenon is the geodetic effect, also known as geodetic precession [GP], which is the product of the GR's spacetime curvature, and was measured close to the Earth as a result of the Earth's mass (Kahn & NASA, 2008). Figure 12, shows the GP in relation to the LTP.

In 1960, Schiff demonstrated that an ideal gyroscope in Earth's orbit, being in the weak−field regime, would experience two relativistic effects: GP and LTP (Schiff, 1960). Schiff's formula is provided in Equation 26. Moreover, the GP moves in the direction of the orbit, being orthogonal to LTP.

$$\boldsymbol{\Omega} = \frac{3GM}{2c^2R^3}(\boldsymbol{R} \times \mathbf{v}) + \frac{GI}{c^2R^3}\left[\frac{3\boldsymbol{R}}{R^2}(\boldsymbol{\omega}.\boldsymbol{R}) - \omega\right] \quad \text{Eq. 26}$$

On the right-hand side of Equation 26, the first term is the GP while the second term is a reworking of Equation 25: LTP, where **R** is the instantaneous distance vector from the Earth, and **v** is the velocity of the gyroscope. GP-B measured the GP to ~0.5% against GR (Kahn & NASA, 2008). Over longer periods of operation, this result was improved to ~ 0.28% (Everitt et al., 2011).

Utilizing Lunar Laser Ranging [LLR][32] as a means of testing GR, provided the value of GP expressed as a deviation to −0.0019 ± 0.0064, with GR being 0 (Williams et al., 2004).

## 2.4 Equivalence Principle tests

The weak equivalence principle [WEP] is the equality of gravitational and inertial mass[33], so that $M_G \equiv M_I$, while Einstein's equivalence principle [EEP] is the equivalence of homogeneous gravity and uniform acceleration in small regions, depicted in Figure 3. However, EEP is dependent on the WEP also being valid (Will, 2014). Tests of the WEP are generally concerned with testing the universality of free fall[34] [UFF]; namely, that all masses in a uniform gravitational field, experience the same acceleration (Merkowitz, 2010).

[32] Apollo 11 astronauts placed a retro-reflector array on the lunar surface for the purpose of LLR.

[33] The first accurate tests of WEP were by Eötvös in 1889 and 1908. The first space-born WEP test was performed by MICROSCOPE (MICROSatellite a traine compensée pour 'Observation du Principe d'Equivalence).

[34] Objects in a free falling, non-rotating, reference frame of a small region becomes the domain of SR.

Some authors dichotomize EEP into two classes, first is EEP and the second is the strong equivalence principle [SEP]. These imply two significant results for locally *non-gravitational* reference frames:

- any experiment is independent of the *velocity* of the UFF reference frame, referred to as local Lorentz invariance [LLI];
- any experiment is independent of *position* and *time* in the universe; this is local position invariance [LPI] (Will, 2014).

If these two statements are valid, then SEP and EEP are correct. Moreover, SEP contains both EEP and the WEP. However, SEP includes the gravitational self-energy of bodies, whereas EEP does not. Hence, SEP requires the involvement of astrophysical objects (Will, 1993). Experiments of EEP have a negligible gravitational self-energy. There have been many tests of the WEP, but this report is only concerned with tests of EEP and the SEP.

Tests of EEP and SEP usually employ the PPN framework and also the Equivalence Principle parameter $\eta$, which is a function of the PPN parameters $\beta$ and $\gamma$, which are often called the Eddington–Robertson–Schiff parameters (Will, 2014). The simplest relationship, using only $\gamma$ and $\beta$, is in Equation 27, often called the Nordtvedt equation; in the weak−field regime.

$$\eta = 4\beta - \gamma - 3 \quad \text{Eq. 27}$$

If EEP and SEP is 100% correct, then $\eta = 0$. Generally, SEP or EEP can be validated by the presence of GvT, or Shapiro time delay, or GvR. Early suggestions of EEP tests were reported in the 1960s (Shamir & Fox, 1969).

In 1979, a hydrogen maser in a space probe at an altitude of 10,000 m was used to measure the GvT, confirmed EEP to $\eta = 2 \times 10^{-4}$ (Vessot & Levine, 1979).

The supernova SN1987A provided a stringent test of EEP on intergalactic scales, by the almost simultaneous arrival of the photons and neutrinos. This result was in agreement with the GR prediction to ~ 0.2% (Longo, 1988). In 2003, data from the Cassini spacecraft, using Shapiro time delay, measured $\gamma = 1.000021 \pm 0.000023$, (Hsu & Fine, 2005; Ni, 2005). Spectral distortions of the Cosmic Microwave Background radiation [CMB], using COBE/FIRAS[35], indicate a deviation from EEP no greater than, $< 10^{-5}$ within CMB energy scales (Arai et al., 2016). EEP has been thoroughly tested (Will, 2014).

[35] FIRAS (Far InfraRed Absolute Spectrophotometer) on board the COBE (COsmic Background Explorer).

Suggestions of various methods to verify the SEP also started in the early 1960s (Morgan & Peres, 1962). LLR has been crucial for testing GR. These tests have provided stringent tests of the SEP and the time variation of *G*. (Merkowitz, 2010).

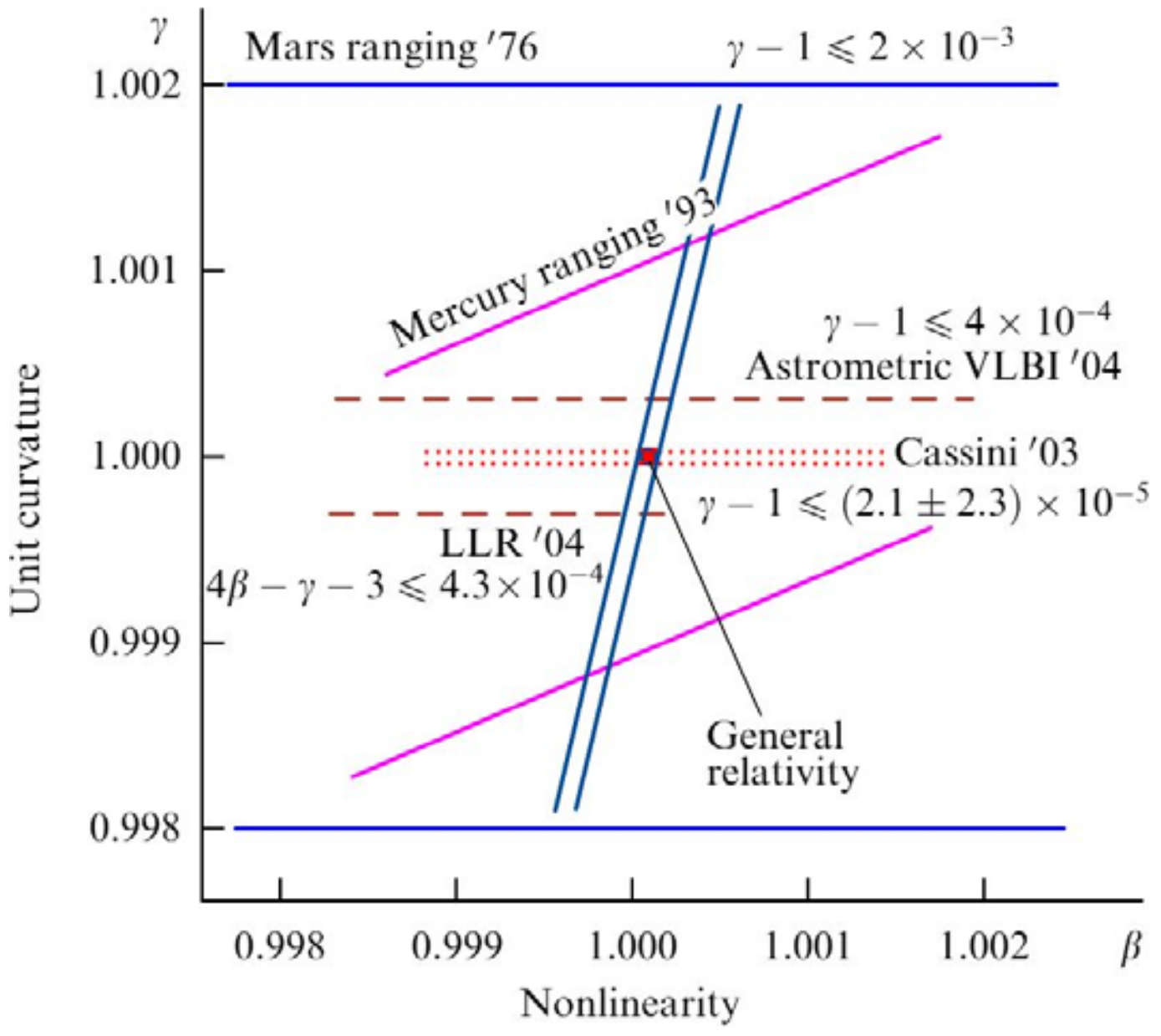

*Figure 13: Values of $\gamma$ plotted against $\beta$ from 1969 to 2009, yielding the best values of $\gamma$–1=(2.1±2.3)×10$^{-5}$ and $\beta$–1=(1.2±1.1)×10$^{-4}$. Courtesy, (Turyshev, 2009).*

LLR researchers were able to test the SEP, using the prediction that gravitational binding energy falls like other forms of mass and energy relations. Over 35 years of LLR tests resulted in $\eta = (2.3 \pm 3.2) \times 10^{-4}$, against the GR prediction of $\eta = 0$ (Adelberger et al., 2009). Recent ground and LLR experiments have yielded a strict upper limit on deviations from the SEP, where $\eta = 4.4 \times 10^{-4}$ (De Marchi & Congedo, 2017). Figure 13, depicts the improvements in determining the Eddington parameters $\gamma$ and $\beta$, within the weak−field regime.

Testing the variation in *time* of fundamental constants[36] are also methods for determining if the SEP is violated. Some details of these methods are provided in Appendix 2.4.1.

Generally, constraining the value of $\eta$ for EEP is a good test for the weak−field regime within the solar system but it can also be employed for the SEP on cosmological scales.

[36] Alternative theories of gravity that violate the SEP predict that *G* may vary with time as the universe evolves.

**2.5 Solar−system tests**

Many of the most important tests of GR have been performed in the solar system, and some have been covered in previous sections of this report. Measurements within the solar system are concerned with the weak−field regime and are considered asymptotically flat in GR. (Blanchet et al., 2011). Moreover, solar system tests have confirmed GR to a high level of accuracy (Ni, 2016). Whilst there have been many solar system studies, and many planned for the future, only a couple notable ones are briefly mentioned.

Nordtvedt, using solar system observations, stated that there was less than 1 part in $10^9$ of a departure from GR (Nordtvedt, 2001). Anderson et al., using the Jet Propulsion Laboratory [JPL] ephemerides data of planetary positions, together with contemporary data analysis methods and PPN parameters, obtained the following. Planetary positions were mapped at different times, it was found that, $(\beta - 1) = -0.0010 \pm 0.0012$, and $(\gamma - 1) = (0 - 0.0015) \pm 0.0021$, whereby the GR predictions are, $(\beta - 1) = (\gamma - 1) = 0$ (Anderson et al., 2002).

EEP and the SEP haven been extensively tested within the weak−field regime.

# Chapter 3:

## Strong−field tests

### 3.1 Gravitational lensing

The use of gravitational lensing, being a result of the bending light, is also considered a test of GR, *primarily* being in the strong−field regime. There are *three* main types of gravitational lensing: 'microlensing,' 'strong lensing' and 'weak lensing.' See Figure 14 for an illustration of gravitational lensing [GL].

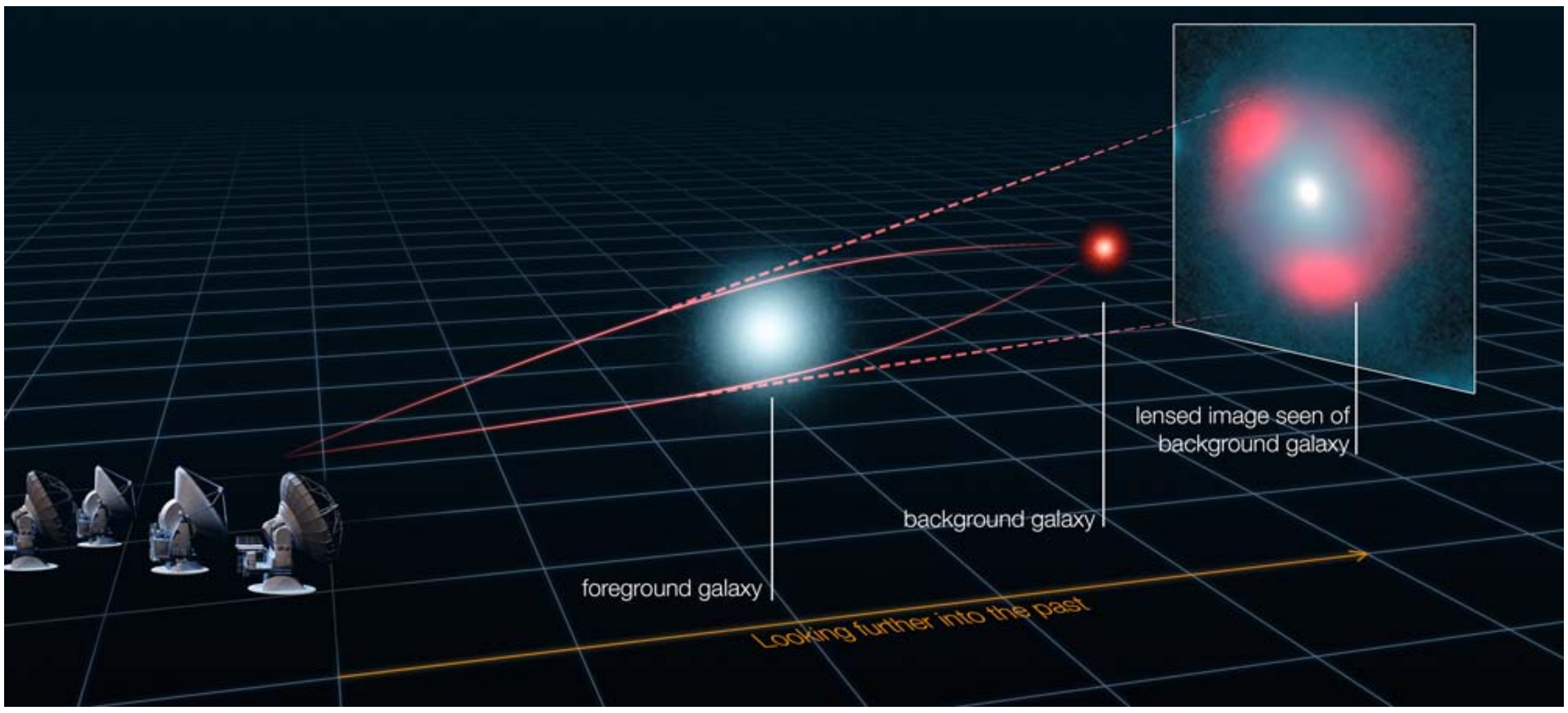

*Figure 14: Illustration of GL. Credit: ALMA (ESO/NRAO/NAOJ), L. Calçada (ESO), Y. Hezaveh et al., http://www.eso.org/public/images/eso1313b/*

The deflection of light in Equation 14 is utilized, with other variables, to produce Figure 15.

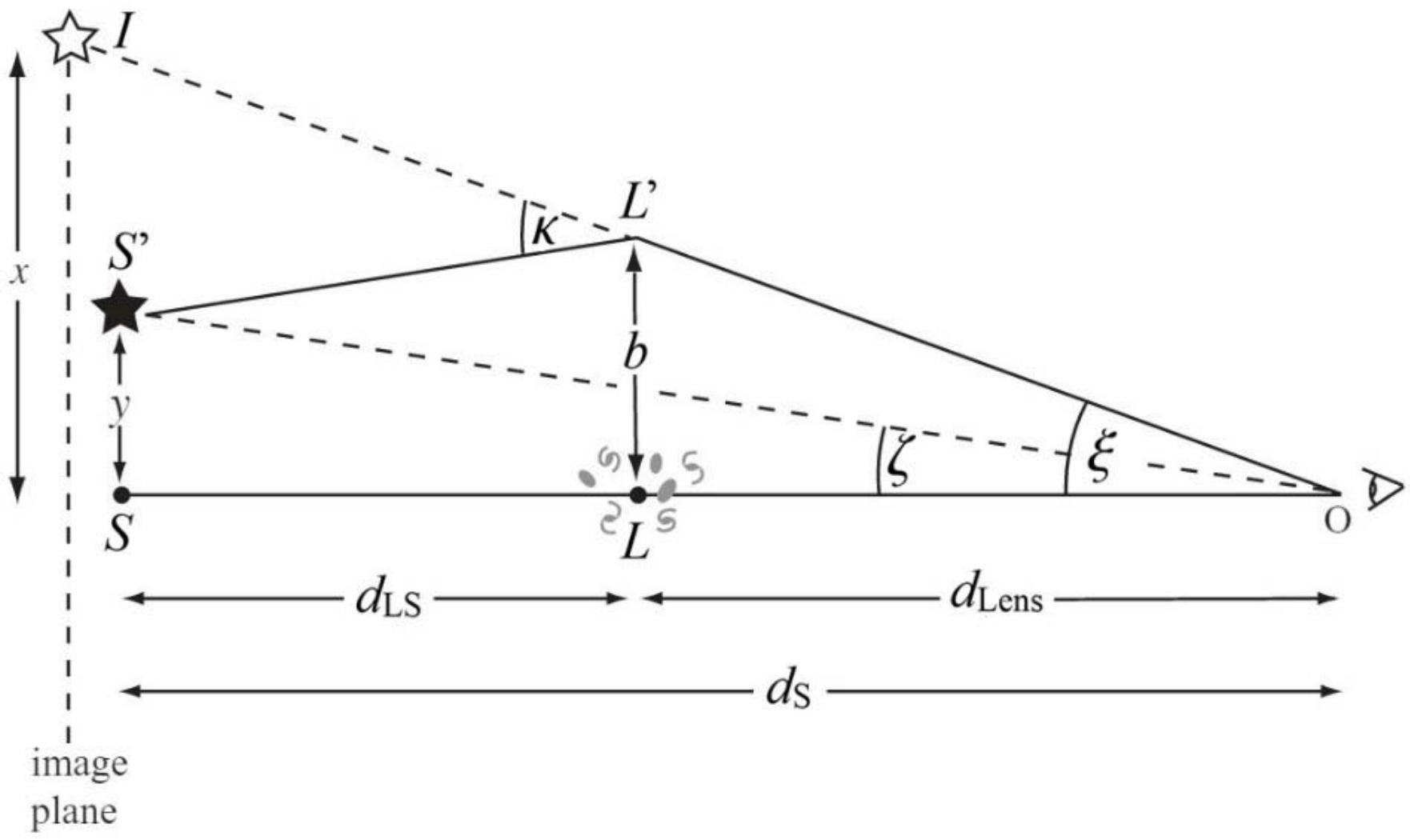

*Figure 15: Lensing illustration. The gravity of the lensing mass M is at L, and bends light from a distant source at S', toward the observer at O. However, the source appears at position I. Image courtesy: (Sparke & Gallagher, 2007).*

The light path is bent by the angle $\kappa$, yet the source appears at an angle $\xi$. Additionally, $\kappa \approx x/d_S$ when $d_S \gg x$. When the bending of light is small; then, the following holds: $x - y = \kappa d_{LS}$. Also, when $d_S \gg b$, then $b = \kappa d_{Lens}$. Using Equation 14, we can represent $\kappa$ as in Equation 28, where $R_S$ is the Schwarzschild radius, as previously shown in Equation 7 (Sparke & Gallagher, 2007).

$$\kappa \approx \frac{4GM}{bc^2} = \frac{2R_S}{b}$$ Eq. 28

This leads to Equation 29, where $\xi_E$ is the 'Einstein radius.'

$$\xi - \zeta = \kappa d_{LS} = \frac{1}{\xi}\frac{4GM}{c^2}\frac{d_{LS}}{d_{Lens}d_S} = \frac{1}{\xi}\xi_E^2$$ Eq. 29

Microlensing is the GL by compact objects along the line−of−sight. The image of the lensed object, often a galaxy, is outside the Einstein radius, being > $\xi_E$. Microlensing magnifies the source, and there is an increase in brightness at the time of closest approach, but the effect may only occur for a limited duration as the lensing object moves in front of the lensed object.

Strong lensing is the most extreme GL and occurs when the lens is very massive, and the source is reasonably close, in distance. In this case, light can take different pathways to an observer producing multiple images. A special case, when the alignment of the lens and source are directly in line−of−sight to the observer, an 'Einstein ring' can be produced around the Einstein radius ~ $\xi_E$, see Figure 15.

Weak lensing is GL of astrophysical objects that are behind an intervening cluster and well outside its Einstein radius, being >$\xi_E$, producing weakly magnified images that are slightly stretched or skewed in tangential directions (Sparke & Gallagher, 2007).

Zhang et al., (2007) determined a method of overcoming galaxy bias[37], by combining measuring large−scale GL, structure growth rate and also galaxy clustering. Together these values produce a quantity $E_G$, referred to as 'gravitational slip.' Different gravitational theories predict differing values of $E_G$. Generally, other theories predict a faster growth rate, producing a smaller value of $E_G$ and so testing for $E_G$ can not only confirm GR but also rule out alternative gravitational theories.

Reyes et al., (2010) calculated the gravitational slip, which took the form of Equation 30.

[37] Galaxy bias: visible matter appears to cluster, within galaxies, more so than dark matter, meaning the spatial clustering of luminous matter does not reflect the total matter present.

$$E_G(R) = \frac{1}{\beta}\frac{\Upsilon_{gm}(R)}{\Upsilon_{gg}(R)}$$

Eq. 30

Where $\beta$ is the 'redshift distortion parameter[38],' so that $\beta \approx \Omega_0^{0.6}/b$ in ΛCMD cosmology, where $\Omega_0$ is cosmological density, and $b$ is the light−to−mass bias (Hamilton, 1998). The values of $\Upsilon_{gm}(R)$ and $\Upsilon_{gg}(R)$ are calculated from observations of large-scale GL and galaxy clustering, respectively. The researchers reported confirmation of GR on large scales, with $E_G = 0.396 \pm 0.06$, against the GR predicted value of, $E_G = 0.4$ (Reyes et al., 2010). Additionally, Alam et al., (2017) also tested GR on large scales using WL together with galaxy clustering data using the Canada–France–Hawaii Lensing Survey [CFHLS] and the Baryon Oscillation Spectroscopic Survey [BOSS] CMASS data. Their determined value was, $E_G = 0.42 \pm 0.056$, against their GR predicted value of, $E_G = 0.396 \pm 0.011$. This result was a 13% deviation from GR, using the Planck 2015 cosmological parameters.

Determination of $E_G$ has confirmed GR and also excluded an alternative theory called tensor–vector–scalar [TeVes][39] gravity theory. See Appendix B.

### 3.2 Gravitational waves

In 1916, Einstein suggested that gravitational waves[40] [GW]s were a firm prediction of GR, being essentially the rippling of spacetime (Einstein, 1916). When the idea of GWs first emerged, "there was ongoing confusion over whether gravitational waves are real or are artefacts of general covariance," (Will, 2014). However, Einstein and Rosen published the first *correct* version of gravitational waves in 1937 (Einstein & Rosen, 1937).

GWs are fundamentally created by mass being accelerated in spacetime. However, if the acceleration is spherically symmetric, then no GWs are radiated. Moreover, binary systems always radiate GWs because their acceleration is asymmetric. The simplest expression for the total amount of gravitational energy or GW luminosity $L_{GW}$, per unit time from a rotating body, is given by the quadrupole[41] in Equation 31. The body's *asymmetry* is denoted by, $\psi$ (Ju et al., 2000). Other variables are: mean radius of the object is $R$, with the rotational velocity of v, with the Schwarzschild radius $R_s$, and a mass of $M$. For a perfect sphere, $\psi = 0$ and no GW is radiated.

[38] 3D redshift maps of galaxies are distorted along the line of sight by peculiar velocities of galaxies.
[39] TeVes is a relativistic generalization of Modified Newtonian dynamics [MOND].
[40] GWs are sometimes also referred to as gravitational radiation.
[41] The gravitational quadrupole represents how stretched out a given mass is, along a specific axis.

$$L_{GW} = \frac{c^5}{G}\psi^2 \left(\frac{R_s}{R}\right)^2 \left(\frac{\mathrm{v}}{c}\right)^6$$ Eq. 31

The ‘*strain*’ $h_s$, is equivalent to the difference between the change in GW luminosity and the GW luminosity, as represented in Equation 32 (Ju et al., 2000).

$$h_s \sim \frac{\Delta L_{GW}}{L_{GW}}$$ Eq. 32

The strain amplitude $h_s$ is related to Equation 32 but takes a different form for various types of GWs. Detecting GWs relies on the detection of small strain amplitudes, while detections may be linear, circumferential, or orthogonal strains. The strain amplitude’s maximum value is proportional to the reciprocal of the distance to the source, in kilometres. For example, the strain amplitude $h_s$, for a black hole at distance $r$, is provided in Equation 33 (Ju et al., 2000).

$$h_s \sim \frac{R_s}{r}$$ Eq. 33

The maximum strain that could be received at the surface of the Earth is $h_s \sim 10^{-21}$ (Hobson et al., 2006). Real detections are far more complex as most GWs are polarised, while GR predicts two degrees of polarisation, other theories predict more, see Appendix 3.2.1, (Lee et al., 2008).

There are four types of GWs. Firstly, there are ‘Continuous GWs,’ which have almost constant frequency and relatively small amplitude, such as binary systems in rotation, or a single object rotating about its axis that may have an extended asymmetric mass. Secondly, are ‘Inspiral GWs,’ which are produced by massive binary systems which are inspiralling towards one another, and as their orbital distances lessens, the rotational velocity increases rapidly. Then there are ‘Burst GWs,’ which are produced by an extreme event such as gamma ray bursters or supernovae. Lastly, there are ‘Stochastic GW,’ which are created in the very early universe and are sometimes called primordial GWs. The sonic waves within the primordial soup would have produced a continuous noise, producing a GW background (Ciufolini, 2001).

The first *indirect* detection of gravitational waves was made in 1974 by Hulse and Taylor of a binary pulsar inspiral PSR 1913 + 16, with an observed orbital decay, using timed radio wave detection (Hulse & Taylor, 1975). The orbital configuration is shown below in Figure 16.

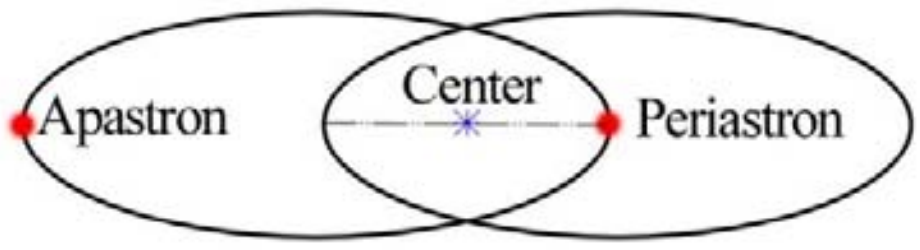

*Figure 16: Binary orbit, with the center being the center of mass. The apastron, being the maximum separation, and periastron or closest approach. Illustration E. Asmodelle.*

Hulse and Taylor found that the gravitational time dilation [GvT] in the binary pulsar system was in agreement with the GR prediction, whereby changes in the pulsars' orbit was consistent with the energy loss due to GWs. The decay of the binary orbit is illustrated in Figure 17, while some Post-Keplerian [PK] parameters are employed. The center of mass orbital period of the pulsar is $P_{\text{cm}}$, while the first derivative is the orbital period decay, $\dot{P}_{\text{cm}}$. The value of $\dot{P}_{\text{cm}}$ against the GR predicted value of $\dot{P}_{\text{GR}}$, was found to be $\dot{P}_{\text{cm}} / \dot{P}_{\text{GR}} = 0.997 \pm 0.002$ (Ashtekar & Petkov, 2014). This is in agreement with GR to better than 1% (Ju et al., 2000).

Moreover, the GW luminosity $L_{GW}$ can be stated as a function of the energy loss over the time of the orbital decay, represented in Equation 34 (Ashtekar & Petkov, 2014).

$$L_{GW} = \left| \frac{dE}{dt} \right| \qquad \text{Eq. 34}$$

Figure 17 shows thirty years' study with PSR 1913+16.

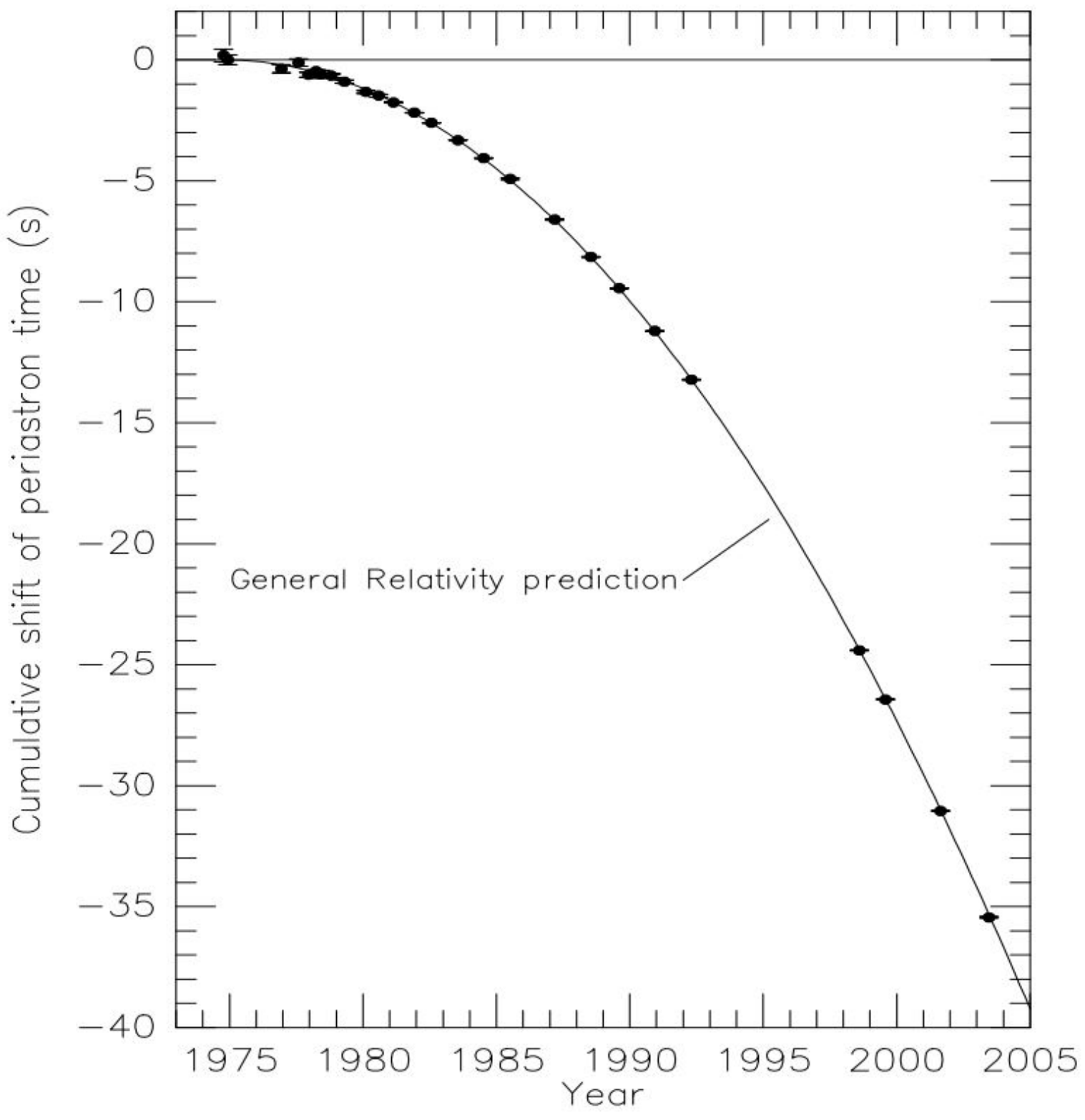

*Figure 17: The orbital decay of PSR B1913+16 or shifting of the periastron. Observations are dots, while the black line is the GR prediction. Courtesy (Weisberg & Taylor, 2004).*

These findings were seminal, "The results all supported general relativity, and most alternative theories of gravity fell by the wayside," (Will, 2014). The Nobel prize, during 1993, was awarded to Hulse and Taylor for their *indirect* discovery of GWs with PSR 1913 + 16.

The first *direct* detection of GWs occurred in 2015 with two different detectors of the Advanced LIGO, from the source GW150914, a black hole binary at a distance of $410^{+160}_{-180}$ Mpc. These corresponded to the coalescing of binary black holes, with initial masses of $36^{+5}_{-4}$ $M_{\odot}$ and $29^{+4}_{-4}$ $M_{\odot}$. The signals were detected in the range ~35 − 250 Hz, and at a strain amplitude of $h_s$ ~$1.0 \times 10^{-21}$ m. The length of the detector arms $L \sim 4 \times 10^3$ m, whereby the two LIGO detectors at Hanford and Livingston received signals over a 16-day period (Abbott, 2016).

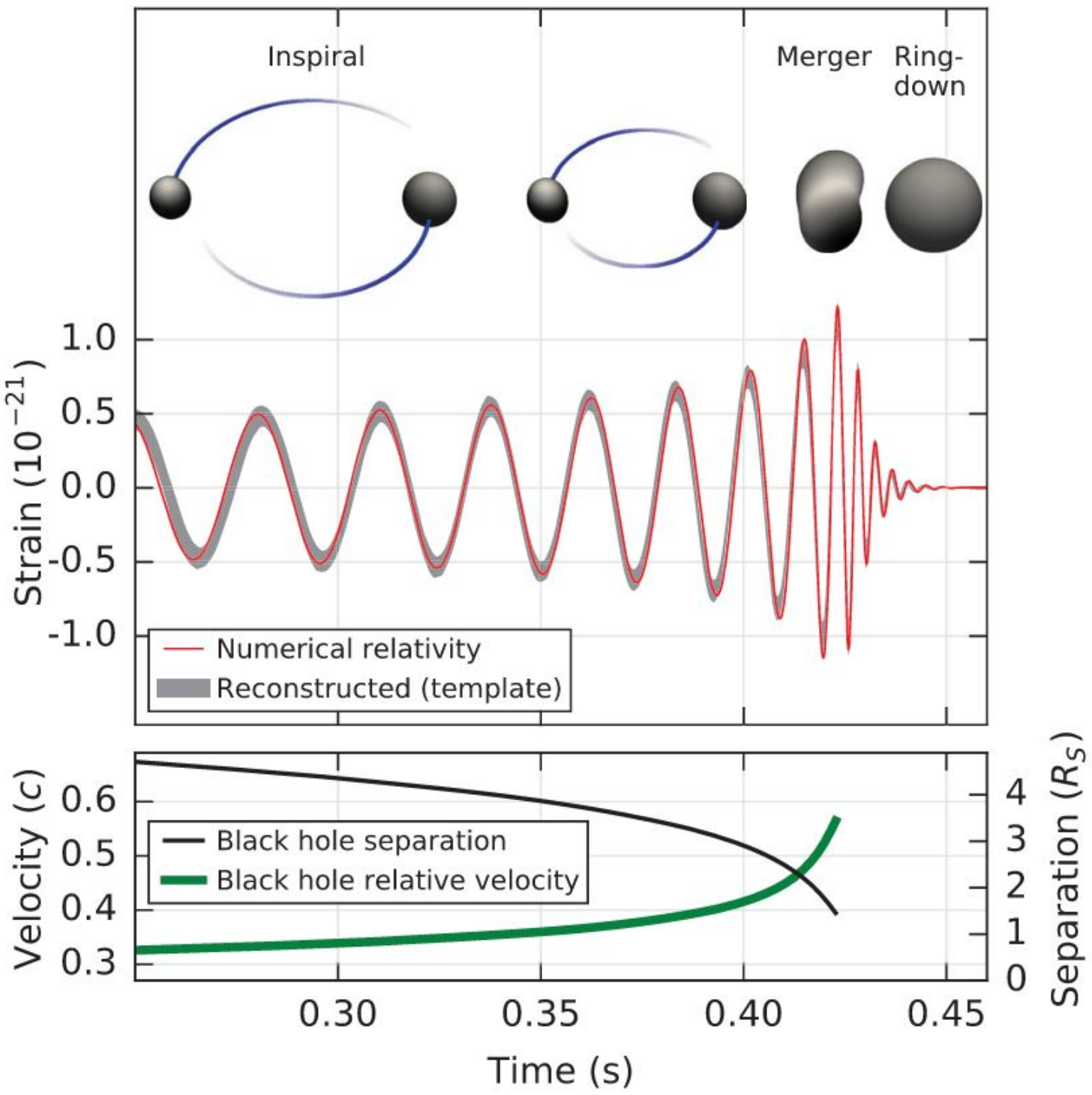

*Figure 18: Top section, shows the calculated waveform using the resulting source parameters from GW150914 reconstructed, compared to simulated waveforms from numerical relativity. The bottom section shows the Keplerian effective black hole separation in units of $R_s$. Credit: LIGO Scientific Collaboration and Virgo Collaboration* (Abbott, 2016)

The closer the black holes were, the more radiation was emitted, which accelerated the inspiralling and produced the characteristic '*chirp*' waveform, whereby both the amplitude and frequency of the GWs increased until peaking at the merger, as depicted in Figure 18.

The detection of GWs from GW150914 is observational verification of spacetime curvature as described by GR. The pre-eminent direct detection of GWs has also opened a new method of observational astronomy and is a monumentally significant result for GR.

### 3.3 Timing pulsars

Timing pulsars, or measuring the rotations of pulsars, can test the combined effects of GvT, and GvR, as well as the Shapiro delay (Shao & Wex, 2016). Additionally, because the gravitational field near pulsars is in the strong−field regime they can also test the SEP, due to the local position invariance [LPI] of objects with strong self−gravitating properties (Will, 1993).

The inspiralling binary pulsar PSR 1913+16 is the first such example. Recently, Weisberg and Huang reported that the ratio between the observed orbital period decay $\dot{P}_{\mathrm{cm}}$, due to energy loss from GWs, and the GR prediction was found to be, 0.9983 ± 0.0016 against unity for GR (Weisberg & Huang, 2016).

Pulsar Timing Arrays [PTA] are a global network of radio telescopes that regularly monitor stable pulsars, and one of the main scientific goals of PTAs is the detection of GWs. Moreover, GWs have a distorting effect on the radio wave pulses that are travelling to a telescope, located on the Earth. A difference can be detected between the actual time−of−arrival [TOA] and expected time of pulses. These differences are called 'timing residuals' $\Delta t$, and they carry crucial GW information. Additionally, the strain $h_s$ is sensitive to the timing residual $\Delta t$, divided by the observation time $T_{\mathrm{obs}}$, as seen in Equation 35 (Mingarelli, 2016).

$$h_s \sim \frac{\Delta t}{T_{\mathrm{obs}}} \qquad \text{Eq. 35}$$

Received radio pulses from a pulsar have their own unique form; to obtain a typical signal profile, many different pulses must be integrated over time, which is referred to as 'phase coherent pulsar timing.' This integration produces a very stable profile with small TOA errors and reveals the pulse dispersion (Mingarelli, 2016). An example is shown above in Figure 19.

The pulsar emission time $t$, and the TOA at the telescope receiver $\tau$, are depicted by Equation 36. The value $D$ is the dispersion delay of the pulsar, in seconds, while the observing frequency $f$, is in MHz. The Einstein delay $\Delta_{\mathrm{E}\odot}$ is comprised of GvT, GvR, and the Shapiro delay $\Delta_{\mathrm{S}\odot}$; both are due to the Sun and solar system objects. The Roemer term $\Delta_{\mathrm{R}\odot}$, consists of the solar system travel time, also proper motion and parallax of the pulsar.

Such calculations are taken as being relative to the Solar System Barycentre [SSB][42].

$$t = \tau - D/f^2 + \Delta_{\mathrm{R}\odot} + \Delta_{\mathrm{E}\odot} - \Delta_{\mathrm{S}\odot} - \Delta_{\mathrm{R}} - \Delta_{\mathrm{E}} - \Delta_{\mathrm{S}} \qquad \text{Eq.36}$$

[42] The Solar System Barycentre [SSB] is considered a quasi-inertial frame, and so the rotational period of a pulsar is nearly constant.

Test results regarding pulsar timing usually incorporate PK parameters, which are functions of the pulsar mass, the companion mass, and the standard five Post-Keplerian [PK] orbital parameters, see Appendix 1.1.5.

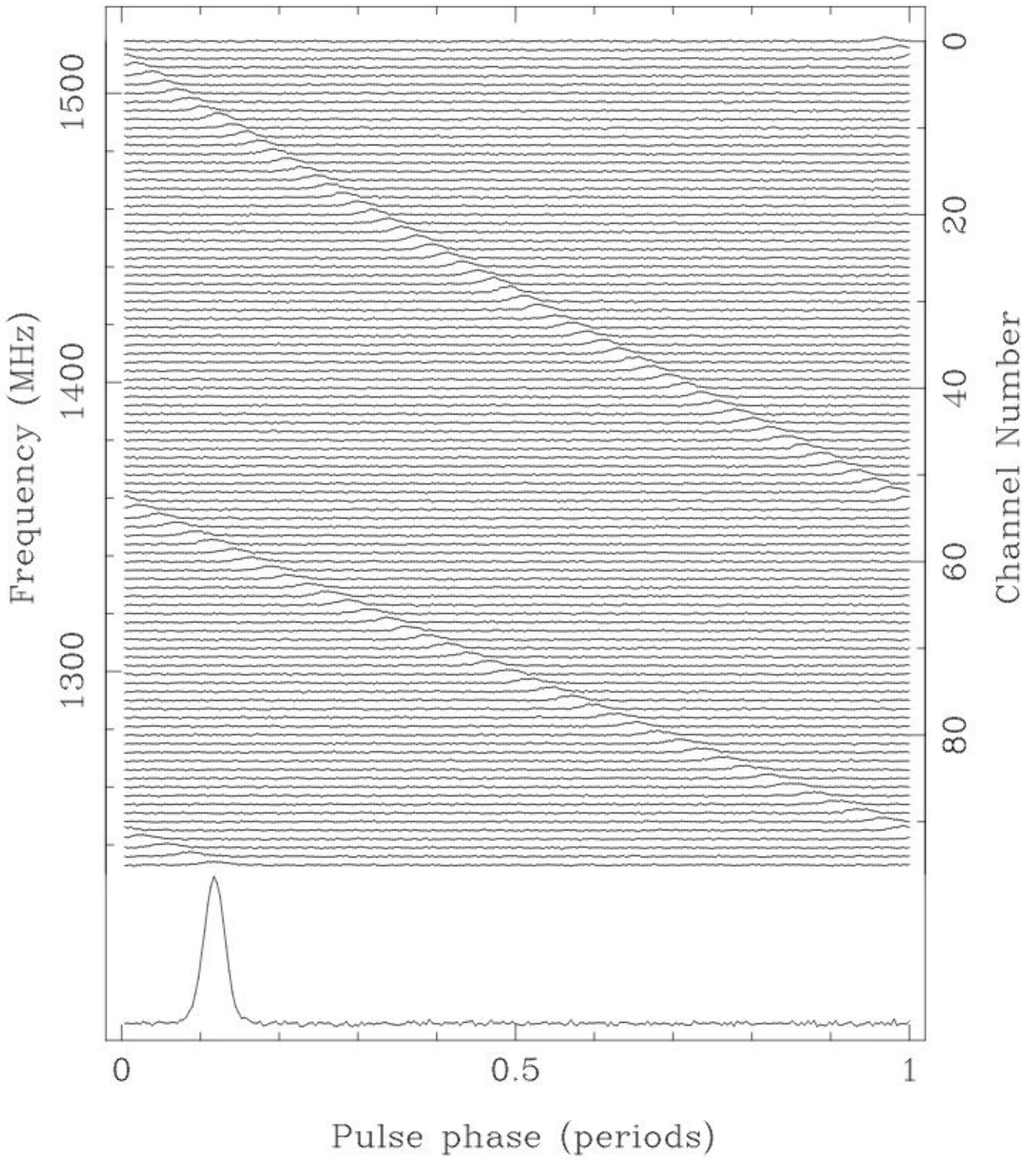

*Figure 19: Diagram shows pulse dispersion of the 128 ms pulsar B1356–60, showing the dispersion delay. The bottom line is an integrated profile. Courtesy (Lorimer, 2008).*

Jacoby et al., (2006) report that the orbital period decay of the double neutron star PSR B2127+11C was determined to be, $\dot{P}_{\rm cm} = (-3.95 \pm 0.13) \times 10^{-12}$, being in line with the predicted GR value to ~3% level. The orbital motion of the double pulsar PSR J0737-3039A/B, using GR corrections, evaluated the Shapiro delay parameter $s$, to be in agreement with the GR predicted value, with an uncertainty of 0.05% (Kramer et al., 2006). These observational tests with PTAs are in the strong−field regime and are considered highly conclusive.

### 3.4 Extreme Environments

The most extreme gravity environments are near very massive compact objects. In these types of environments, spacetime curvature is highly pronounced and general relativistic effects are profound. Such massive objects are generally neutron stars and black holes [BH]s, while the most extreme version of the BH is the super−massive black hole [SMBH]. Active galactic nuclei [AGNs] and quasars hold SMBHs as their central engine and are also suitable general relativistic test laboratories. Moreover, any deviation in observed phenomena with GR will be most prevalent in these environments, being in the strong−field regime.

BH tests of GR can probe the strong−field regime, while many of these tests have yet to be performed. One such test, comprising 16 years of observations, was carried out by Gillessen et al., (2009) of Sagittarius A* [Sgr A*], referred to as Sagittarius A-star. Sgr A* is a bright radio source in the center of the Milky Way, which is driven by a SMBH. The researchers imaged the orbits of stars around Sgr A*, called S-stars.

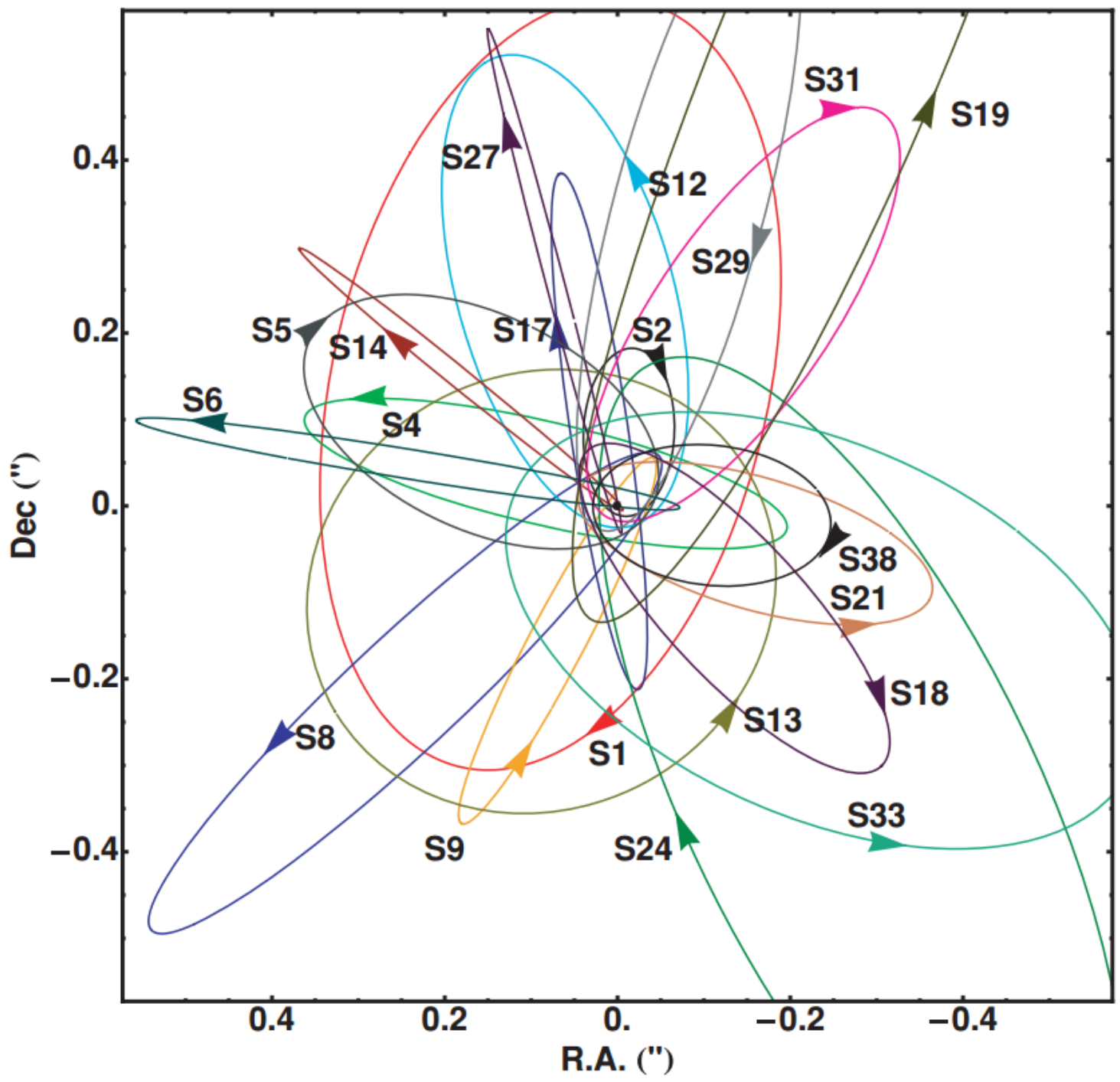

*Figure 20:Orbits of S-stars around the central SMBH. The coordinate system chosen is so Sgr A* is at rest (Gillessen et al., 2009).*

Figure 20 shows the orbits of the S-stars. Gillessen et al., (2009) used a Plummer[43] profile for fitting, taking into account relativistic effects and obtained the following Nordtvedt value for the upper limit of the SMBH: $\eta = 0.021 \pm 0.019$, against the GR value of 0.

Hambaryan et al., (2017) through observation and measurement determined the GvR of the neutron star, RX J0720.43125, to be $z_g = 0.205^{+0.006}_{-0.003}$, being completely consistent with GR.

[43] A Plummer profile is a density relation which has a finite-density core and falls off as $r^{-5}$ which is much steeper fall than is seen generally in most galaxies.

Quasi-stellar objects [QSOs], or simply quasars, are the most luminous in the AGN classes. Moreover, QSOs are considered the brightest objects in the Universe, except for the short-lived explosions related to supernovae and gamma-ray bursters. The majority of QSOs are radio sources. However, the unified model of AGNs explains both radio-loud and radio-quite QSOs, Seyfert[44] galaxies, and Blazers[45], with a reasonable explanation for the sources of the narrow-line region [NLR] and the broad-line region [BLR] of AGNs. In actual fact, there is a zoo of such objects that are covered by this model; see Figure 21 for the basic ones mentioned.

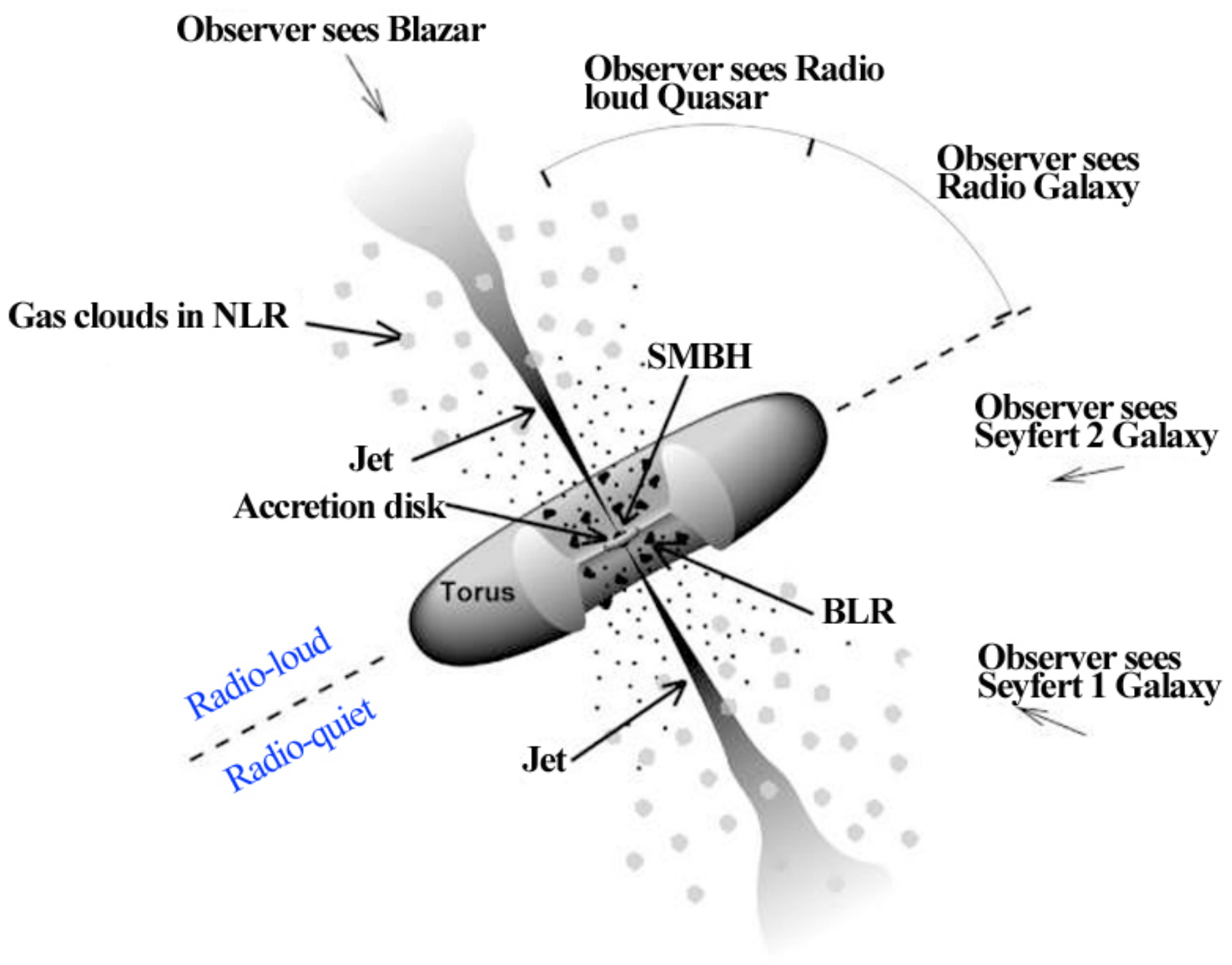

*Figure 21: The AGN unified model, showing the BLR and NLR, as well as different types based on the observational angle. Credit: Fermi and NASA: https://fermi.gsfc.nasa.gov/science/eteu/agn/*

QSO spectra are characterised by a relatively flat continuum with broad emissions, while broadened QSO spectra originate in the BLR close to the SMBH, see Figure 22. The most notable are the Lyman $\alpha$, and ~90% of QSOs are broad-line emitters (D'Onofrio et al., 2012).

Generally, the broad emission lines of QSOs are permitted while the narrow emission lines can be permitted and forbidden. The broadened lines, emanating from the BLR, are evidence of general relativistic effects, by way of GvR. Figure 22 shows the GR component of the line profile. QSO spectra also reveal absorption lines of the intervening gas (D'Onofrio et al., 2012).

[44] Seyfert galaxies type 1, show both narrow and broadened optical emission lines, while type 2 have only narrow emission lines which are wider than such lines in normal galaxies.

[45] Blazars: a class of AGN, being radio sources; are highly variable and do not show emission lines in their spectra.

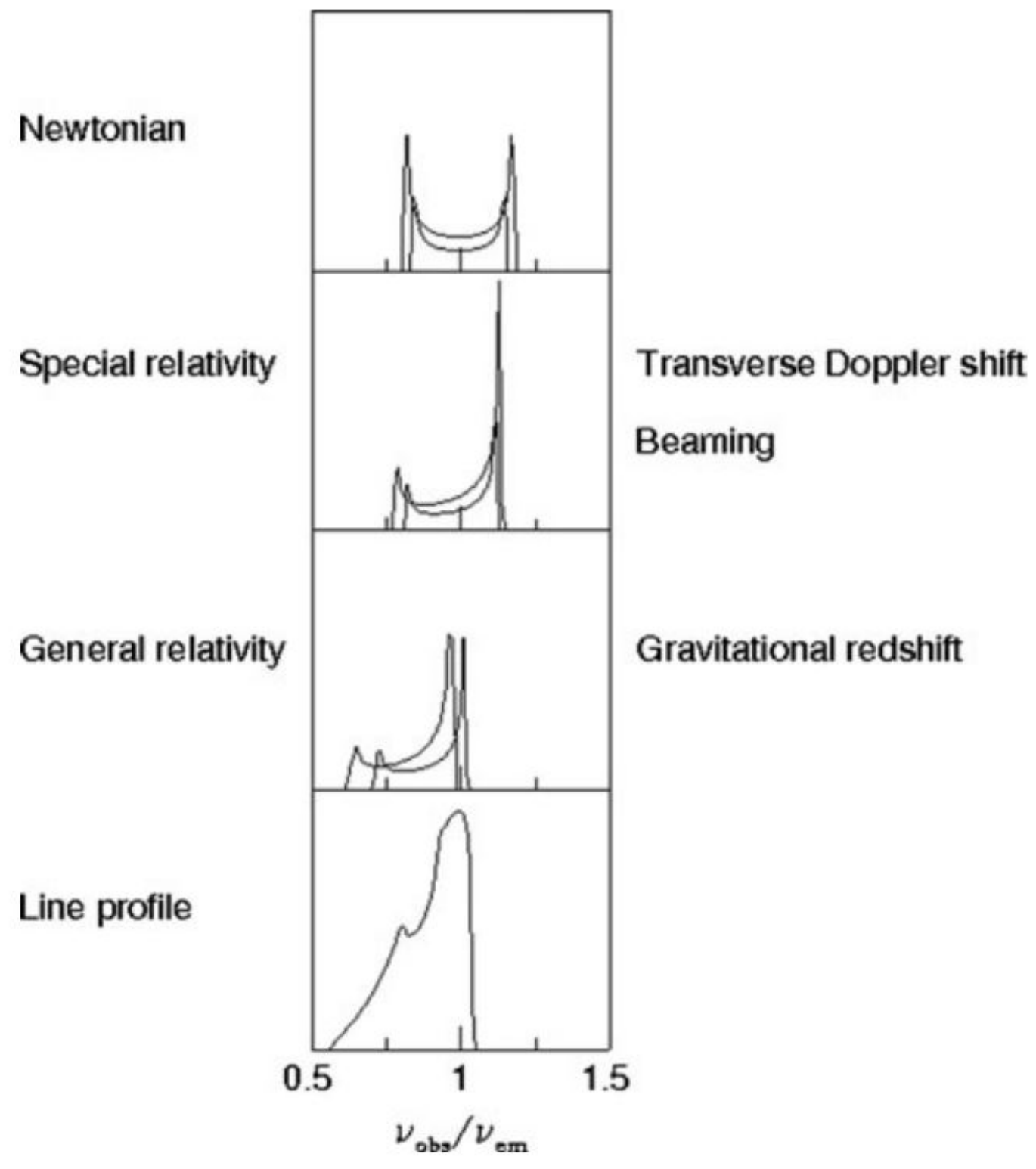

*Figure 22: The components of QSO spectra are shown, Newtonian, special relativistic, and general relativistic. The line profile at the bottom is the combination of all those components. Credit: (D'Onofrio et al., 2012).*

QSO redshifts have been the source of much controversy and reporting ever since their discovery in the late 1950s. However, is generally accepted that broadening of absorption lines from the BLR are evidently the result of GvR. More recently, Tremaine et al., have determined the redshift of the broad line H$\beta$, for 20,000 quasars and found that mean redshift is predominately due to general relativistic effects, in particular, GvR (Tremaine et al., 2014).

Confirmation of GvR in the strong−field regime with QSOs has been instrumental in confirming GR and also constraining the unified AGN model.

# Chapter 4:

## Cosmological tests

### 4.1 Expanding universe

In 1922, Friedmann derived an exact solution of the field equations of GR, describing an isotropic and homogeneous universe. However, the Friedmann solution suggested that such a universe should *either* expand or contract, which was contrary to the belief, at the time, that the universe was static. Einstein then added the cosmological constant Λ to his field equations to make such a dynamic universe, become static (Eigenbrod, 2011). Then in 1927, Lemaître quite independently derived equations similar to that of Friedmann, but Lemaître suggested the universe was *expanding* from the point of 'creation.' Observations by Hubble et al., in 1927 and 1929, proved that the universe was expanding (Das & DeBenedictis, 2012). The Friedmann−Lemaître equations, incorporating Λ, are shown in Equations 37 and 38.

$$\frac{\ddot{a}}{a} = -\frac{4\pi G}{3}(\rho + 3\rho) + \frac{\Lambda}{3} \qquad \text{Eq. 37}$$

$$\left(\frac{\dot{a}}{a}\right)^2 = \frac{8\pi G}{3}\rho + \frac{kc^2}{a^2} + \frac{\Lambda}{3} \qquad \text{Eq. 38}$$

The scale factor is *a(t)* so that in the present epoch *a(t)* = 1. The curvature parameter is *k*, which is time-independent. The total density is, $\rho = \rho_m + \rho_{rad} + \rho_\Lambda$, where the total is the sum of densities of matter, radiation and Λ, or vacuum, respectively (Eigenbrod, 2011).

As a self-test, the general theory of relativity predicts the expanding universe, via solutions to the field equations by Friedmann and Lemaître. Hence, the Friedmann−Lemaître equations are a theoretical proof of GR; confirmed by observation.

### 4.2 Cosmological observations

GR has successfully been implemented to calculate the dynamics of spacetime and the distribution of the large-scale structure of galaxies and radiation (Bean et al., 2011). Furthermore, "tests of general relativity are still in their infancy on cosmological scales," (Bull, 2016).

The detection of the CMB is a prediction of big bang cosmology [BBC], of which GR is fundamental. BBC suggests that a perfect blackbody spectrum should be detected, while the observed value of $T_{CMB}$ = 2.725 K fits very well (Bucher & Ni, 2015). More specifically, within Lambda cold dark matter [ΛCMD] cosmology, "general relativity theory passes tests to some fifteen orders of magnitude from length scales of the precision tests of gravity physics," (Peebles, 2005). Direct tests of GR on cosmological scales, have to take into account a non-zero Λ, dark energy or accelerated expansion, and dark matter. Moreover, there are many ways to test GR cosmologically, but only a couple of methods are explored.

Measuring the observed growth rate of galaxy clusters is one method. Rapetti et al. (2010), measured observations of 238 clusters from the ROSAT All-Sky Survey and data from the Chandra X-ray Observatory. They measured the growth index $\gamma_G$, using a spatially flat model, with given assumptions and including Λ, found $\gamma_G (\sigma_8 / 0.8)^{6.8} = 0.55^{+0.13}_{-0.10}$ , whereby the values for $\sigma_8$ range from 0.79–0.89. The GR value prediction was, $\gamma_G \sim 0.55$. The researchers found no deviation with GR in ΛCMD cosmology (Rapetti et al., 2010).

Measuring the temporal variability of constants over the life of the universe is another method. Will (2014) states that deviations from GR on a cosmological scale range from, $10^{-5}$ to $10^{-7}$, at most, where $\gamma = 1$ in GR, and $|\gamma - 1| = 0$. If local position invariance [LPI] holds; then, the fundamental non−gravitational constants would not have varied in time, yet some researchers believe they may have (Webb et al., 1999; Webb et al., 2011), see Appendix 2.4.1.

In contrast, Kotuš et al. (2017), constrained a limit on the fine structure constant $\alpha$, using the bright QSO HE0515-4414. They found the temporal change of $\alpha$ to be $\Delta\alpha/\alpha = -1.42 \pm 0.55_{stat} \pm 0.65_{sys}$ ppm, which is consistent with *no* temporal change to $\alpha$ (Kotuš et al., 2017).

Additionally, testable violations of the SEP could occur within the WEP with a varying gravitational constant *G*. If the WEP was violated, then the SEP would be violated, and GR would be invalid. An upper constraint of $\Delta G = 4 \times 10^{-20}$ was measured in the orbits of binary millisecond pulsars, more specifically in their anomalous eccentricities. This result shows *no* temporal variation in *G*. Moreover, the variation should be of the order of the expansion, so that, $\dot{G}/G \sim H_0$, whereby $H_0$ is the Hubble expansion parameter (Will, 2014).

### 4.3 Weak lensing surveys

Weak lensing [WL] surveys are also in their infancy, but some surveys have already obtained compelling results. WL produce distortions in the apparent image of lensed astrophysical object's size, shape, and fluxes. The magnification and shearing of high-redshift sources, due to the presence of fluctuations in the intervening gravitational potential, is a useful tool to test

GR. Not only is WL a good method for testing GR, but it is also a powerful probe of dark energy and dark matter (Song & Doré, 2009).

The work of Reyes et al., with WL surveys, was previously discussed in section 3.1 (Reyes et al., 2010) and is still one of the best tests of GR using WL. They measured the 'gravitational slip,' being the difference between the two different gravitational potentials,[46] which define the matter perturbations in the gravity metric. In GR this value is zero, or slight, but in other theories, it is non-zero and leads to substantial differences in the growth of structures and the strength of GL (Reyes et al., 2010). "In GR, the slip is generated by the relativistic shear stress and is considered negligible in the matter-dominated era," (Bertschinger, 2011). The total potential $\Phi_{len}$ of the lens is shown in Equation 38, where $\Phi$ is the Newtonian potential and $\Psi$ is the second gravitational potential.

$$\Phi_{\text{len}} = \frac{\Psi - \Phi}{\Phi} \qquad \text{Eq. 39}$$

Daniel & Linder (2010) compared CMB data from CFHLS with the COSMOS[47] weak lensing data and found *no* deviation from GR, to 99% confidence level limit at $z < 1$. Song et al., (2011) similarity used a combination of galaxy peculiar velocity measurements and WL to constrain modifications to GR. They found *no* deviation from GT on cosmological scales.

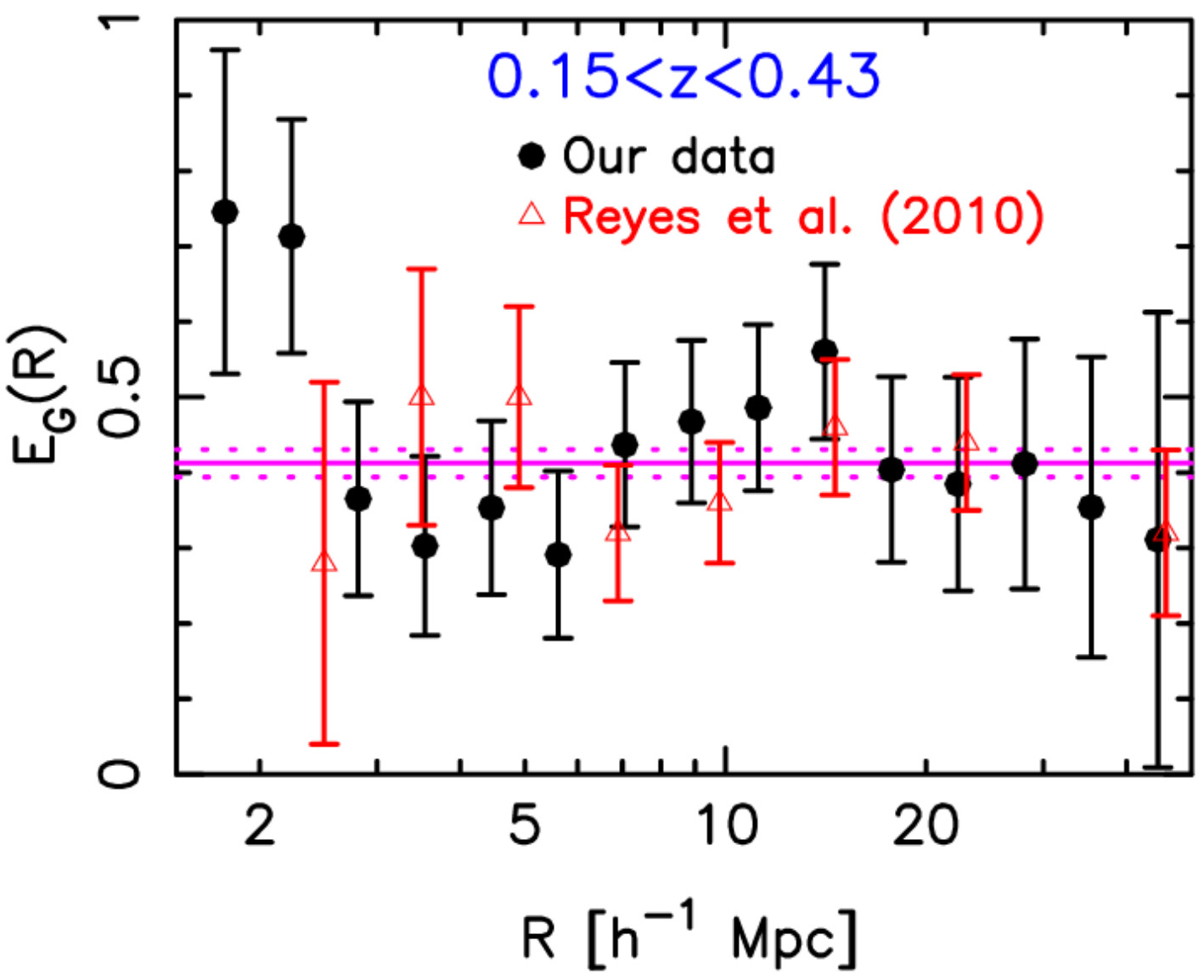

*Figure 23: $E_G$ measurements against radius R, at 0.15<z<0.43 and 0.43<z<0.7. The measurements of Reyes et al. are in red, while Blake et al., data is in black. The horizontal pink solid line is the GR prediction. Courtesy: (Blake et al., 2016)*

[46] The 'gravitational slip' is the difference from two gravitational potentials. Alternative theories of gravity, predict this value could be relatively high.

[47] COSMOS is a weak gravitational lensing shear catalogue

More recently, Blake et al. (2016), performed similar GR tests on cosmological distances, using spectroscopic data and imaging from the WiggleZ Dark Energy Survey, Baryon Oscillation Spectroscopic Survey, Red Cluster Sequence Lensing Survey, and the CFHLS. They found that the gravitational slip statistic $E_G = 0.48 \pm 0.10$ at $z = 0.32$ and also $E_G = 0.30 \pm 0.07$ at $z = 0.57$, while the GR predicted values were, $E_G = 0.41$ and 0.36 for respective redshifts. This validates GR at those redshifts. See Figure 23 for a plot of data.

The enormous amount of confirmational test data is compelling, yet General relativity continues to be tested in future missions, as can be seen in Appendix 4.3.

# Conclusion

Einstein's equivalence principle [EEP] was the fundamental basis for the successful development of the general theory of relativity [GR] and so testing EEP and the strong equivalence principle [SEP] to the highest order validates GR and stipulates that gravity must necessarily be curved spacetime.

Einstein showed that gravitational time dilation [GvT] is the causal mechanism of the gravitational redshift [GvR], in so far as the frequency of light as it climbs out of a gravitational potential becomes redshifted as it undergoes time dilation (Einstein, 1907). Additionally, GvT is also the causal effect of the bending of light and also Shapiro time delay. Hence, indirect confirmations of GvT are the experimental and observational confirmations of GvR.

Einstein's accurate calculations of the precession of the perihelion of Mercury was the *first* verification of the general theory of relativity (Harper, 2007), followed by Eddington's observations of the bending of starlight.

Gravitational lensing [GL] has clearly become a paramount tool in astronomy and cosmology, having a huge variety of applications. Moreover, GL is being used routinely and is a recurrent validation of GR, in the strong−field regime.

The supra−nuclear densities and extreme compactness of pulsars provide an excellent laboratory for testing GR. Pulsar rotations, especially rotations of millisecond pulsars, are very stable and can be considered as almost perfect Einstein clocks (van Haasteren, 2014). The direct detection of gravitational waves [GW]s has opened a new era in astronomy, with a huge potential to determine more accurately the GR constraints for coalescing binary systems.

It was Wheeler who first coined the phrase, 'black hole' [BH] (Schutz, 2003), yet Einstein was not convinced of their existence! However, GR directly implies their existence, and with more observations and measurements, the details of BHs will unfold. So too, the presence of GvR in quasars [QSOs] still requires more study, while the very existence of GvR with QSOs is in a practical sense a confirmational test of GR.

WL surveys will allow enhanced mapping of matter perturbations and their associated gravitational potentials. This will also test the cosmological principle and probe dark matter and dark energy at higher levels (Zhao et al., 2009). Forthcoming WL surveys promise greater precision and verification of GR at larger cosmological scales.

So, should GR continue to be tested? The answer is a resounding, yes! With the concordance ΛCDM model of the universe, cosmological tests of Einstein's theory of general relativity are only just beginning. Furthermore, within both the weak−field and strong−field regimes, PPN values need to be constrained to the point where they are equivalent to GR's predictions.

Many researchers are considering alternative theories of gravity, yet GR testing continues to higher levels of precision. Some even consider modifying GR to suit observation. However, it is understood that "no deviations from Einstein Theory of Gravity have been detected so far from observations of neutron stars," (Yakovlev, 2016). Moreover, "General relativity has held up under extensive experimental scrutiny," (Will, 2014).

So, what is the future of GR? Well continued tests of GR on the solar system and cosmological scales will certainly give us a clear indication if the general theory of relativity is fine the way it is or if some modification is required.

## Acknowledgements

I would like to acknowledge the enduring support and advice from the course leader Dr. Barbara Hassall during recent years and Dr. Silvia Dalla for her guidance in previous years. I would also like to thank my dissertation supervisor Dr. Tim Cawthorne for his analytical guidance this past distance year. Generally, I wish also to acknowledge the University of Central Lancashire, in particular, the School of Computing, Engineering and Physical Sciences for offering the astronomy distance program, may it continue to flourish.

This dissertation would not have been possible without extensive use of SAO/NASA's Astrophysics Data System run by Harvard-Smithsonian Center for Astrophysics, the pre-print service ArXiv, run by Cornell University, and the UCLAN on-line Library.

# *Appendices*

**Appendix A: Acronyms**

| | |
|---|---|
| AGN | Active galactic nuclei |
| BBC | Big bang cosmology |
| BHB | Black hole Binaries |
| BH | Black hole |
| BLR | Broad-line region |
| CMB | Cosmic Microwave Background radiation |
| DES | Dark Energy Survey |
| EEP | Einstein's equivalence principle |
| ESA | European Space Agency |
| GP | Geodetic precession |
| GPS | Global Positioning System |
| GL | Gravitational lensing |
| GvR | Gravitational redshift |
| GvT | Gravitational time dilation |
| GW | Gravitational Wave |
| GP-A | Gravity Probe A |
| HST | Hubble Space Telescope |
| JPL | Jet Propulsion Laboratory |
| LISA | Laser Interferometer Space Antenna |
| LTP | Lense–Thirring precession |
| LLI | Local Lorentz invariance |
| LPI | Local position invariance |
| LLR | Lunar Laser Ranging |
| MBHB | Massive black hole Binaries |
| MOND | Modified Newtonian Dynamics |
| NLR | Narrow-line region |
| PPN | Parameterized post-Newtonian |
| PK | Post-Keplerian |
| PTA | Pulsar Timing Arrays |
| QSO | Quasi-stellar object or quasar |
| Sgr A* | Sagittarius A* |
| SSB | Solar system Barycentre |
| SR | Special relativity |
| SKA | Square Kilometre Array |
| SEP | Strong equivalence principle |
| SMBH | Super-massive black hole |
| TeVes | Tensor-Vector-Scalar |
| LSST | The Large Synoptic Survey Telescope |
| TOA | Time-of-arrival |
| USNO | U.S. Naval Observatory |
| UTC | Universal Coordinated Time |
| UFF | Universality of free fall |
| VLBI | Very-long baseline interferometry |
| WL | Weak lensing |

## Appendix B: Gravitational Theories

A comparison of some alternative theories to General Relativity are shown in Table 1.

| Theory | Metric | Other Fields | Free Elements | Status |
|---|---|---|---|---|
| Newton 1687 | Non-metric | P | None | Nonrelativistic |
| Poincaré 1890-1900 | – | – | – | Fails |
| Nordström | Minkowski | S | None | Fails |
| Einstein [GR] 1915 | Dy | None | None | Viable |
| Whitehead 1922 | – | – | – | Violates LLI |
| Cartan 1922-1925 | ST | – | – | Still viable |
| Kaluza-Klein 1920 | T | S | Extra dimensions | Violates SEP |
| Birkhoff 1943 | T | – | – | Fails |
| Milne 1948 | Machain BG | – | – | Incomplete |
| Thiry 1948 | ST | – | – | Unlikely |
| Belifante-Swihart 1957 | Non-metric | T | K param | Violates EEP |
| Brans-Dicke 1961 | Generic S | Dy | S | Viable for $\omega > 500$ |
| Ni 1972 | Minkowski | T, V, S | 1 param, 3 ftns | Violates LPI |
| Will-Nordtvedt | Dy T | V | – | Viable at high energies |
| Barker 1978 | ST | – | – | Unlikely |
| Rosen 1973 | Fixed | T | None | Violated by pulsars |
| Rastall 1976 | Minkowski | S, V | None | Violated by GWs |
| $f(R)$ models 1970s | $n$ + 1ST | S | Free ftn | Viable but constrained |
| MOND 1983 | Non-metric | P | Free ftn | Viable - Nonrelativistic |
| DGP 2000 | ST/Quantum | – | – | Contradicts BAOs & CMB |
| TeVeS | T, V, S | Dy S | Free ftn | Highly unstable |

*Table 1: List of gravitational theories, excluding the 2016 Emergent Gravity. Abbreviations: Tensor: T, Vector: V, Scalar: S, Potential: P, (Dynamic: Dy, Parameter: param, Function: ftn. Credit: (Debono & Smoot, 2016)*

**Appendix 1.1.1: Newtonian gravitational potential**

The Newtonian gravitational potential also obeys the Poisson equation, see Equation 40.

$$\nabla^2\Phi = 4\pi G\rho \qquad \text{Eq. 40}$$

The Newtonian gravitational potential $\Phi(x, t)$ obeys Poisson's equation $\nabla^2\Phi = 4\pi G\rho$, where $G$ is Newton's constant, and $\rho(x, t)$ is the density of gravitational mass that is active.

**Appendix 1.1.2: Einstein's field equations**

A breakthrough came in 1913 while Einstein was working with Marcel Grossmann (Einstein & Grossmann 1913), in which the Newtonian scalar was replaced by the metric tensor $g_{\mu\nu}$ which determines the curvature of spacetime. The four coordinates[48] $(x, y, z, ct)$ are rewritten $(x_1, x_2, x_3, x_4)$.

The spacetime interval was defined as shown below, in its original form. This indicates that objects move through spacetime along geodesics[49], where $ds$ is the line element.

$$ds^2 = \sum_{\mu\nu} g_{\mu\nu} dx_\mu dx_\nu \qquad \text{Eq. 41}$$

On the 25th of November 1915, Einstein finally published the correct version of the generally covariant[50] field equations of general relativity [GR], using the principle of general covariance[51] (Einstein 1915a). This also implied the principle of consistency[52]. The essence, in tensor form[53], states that the motion of matter is determined in terms of energy and momentum.

A modern version of the field equations is shown below, with the cosmological constant[54], Λ (Schutz 2011). Einstein introduced Λ to realise, what was then considered, a static universe and to fully implement Mach's principle, yet it eluded full implementation (Einstein 2013; Hsu & Fine 2005).

---

[48] *ct* is the speed of light multiplied by time, calibrating c in terms of length units.

[49] Geodesics are extremal paths, where the interval between two points is, either a maximum or a minimum.

[50] A covariant transformation specifies how vectors or tensors, change under a change of basis, while the inverse is a contravariant transformation. It is possible to transform from covariant to contravariant and vice versa.

[51] The principle of general covariance: states that the laws of physics should take the same form in all reference frames; expressed as tensor relationships that are covariant under general coordinate transformations.

[52]The principle of consistency: a new theory should include the successful predictions of earlier theories.

[53] Later Einstein developed a convention, whereby the sigma notation was dropped as it was implied.

[54] Einstein added Λ later, and then removed it when Hubble determined the universe was expanding in 1929. More recently, with accelerated cosmic expansion, it has been reintroduced.

$$G^{\alpha\beta} + \Lambda g^{\alpha\beta} \equiv kT^{\alpha\beta}$$ Eq. 42

There are two constants, Λ and $k$, while $k = 8\pi G/c^4$ (when geometrised, $G/c^4 = 1$). $G^{\alpha\beta}$ is the Einstein tensor, which is symmetric. While $g^{\alpha\beta}$ are elements of the metric tensor, and $T^{\alpha\beta}$ is the energy–momentum tensor. Equation 42, represents 16 partial differential equations, while 6 are duplicates, leaving 10 remaining (Lambourne 2010).

### Appendix 1.1.3: Schwarzschild line element

A modern version of the Schwarzschild line element is provided in Equation 43 (Schutz 2011).

$$\mathrm{d}s^2 = -\left[1 - \frac{2M}{r}\right]\mathrm{d}t^2 + \left[1 - \frac{2M}{r}\right]^{-1}\mathrm{d}r^2 + r^2\mathrm{d}\Omega^2$$ Eq. 43

This is an exterior metric of a gravitating mass $M$, while $t$ is coordinate time, $r$ is the exterior distance from the centre of that body, and Ω is the density parameter. Many other solutions would follow.

### Appendix 1.1.4: PPN values

See Table 2 for details of all PPN parameters, with their values in GR. The two PPN parameters that are used to test GR; namely, $\gamma$ and $\beta$ are equal to unity, and they are often used as follows $(\gamma - 1)$ or $(\beta - 1)$, such that for a 100% GR validation they should be zero or close to it. See Table 3 for constraints on $\gamma$ and $\beta$.

| Parameter | Description in GR | Value in GR |
|---|---|---|
| $\gamma$ | spacetime curvature produced, per unit rest mass | 1 |
| $\beta$ | nonlinearity of gravity | 1 |
| $\xi$ | preferred location effects | 0 |
| $\alpha_1$ | preferred frame effects | 0 |
| $\alpha_2$ | | 0 |
| $\alpha_3$ | | 0 |
| $\alpha_3$ | violation of conservation of momentum | 0 |
| $\zeta_1$ | | 0 |
| $\zeta_2$ | | 0 |
| $\zeta_3$ | | 0 |
| $\zeta_4$ | | 0 |

*Table 2: The parameter $\alpha_3$ is shown twice as it describes two effects. Table courtesy, (Kopeikin et al. 2011)*

| *Parameter* | *Effect* | *Limit* |
|---|---|---|
| $\gamma - 1$ | Time delay | $2.3 \times 10^{-5}$ |
| | Light deflection | $2 \times 10^{-4}$ |
| $\beta - 1$ | Perihelion shift | $8 \times 10^{-5}$ |
| | Nordtvedt effect | $2.3 \times 10^{-5}$ |

*Table 3: Current limits on the PPN parameters.*
*Table data of parameterized elements (Will 2014).*

**Appendix 1.1.5: Post-Keplerian [PK] parameters for GR**

The Post-Keplerian [PK] parameters are shown in Table 5.

| **Post-Keplerian [PK] parameters** | |
|---|---|
| Longitude of periastron | $\omega$ |
| Advance of periastron | $\dot{\omega}$ |
| Orbital period | $P_b$ |
| Orbital period derivative | $\dot{P}_b$ |
| Gravitational redshift [GvR] | $\gamma$ |
| Eccentricity | $e$ |
| Range of Shapiro delay | $r$ |
| Shape of Shapiro delay | $s$ |
| Angle of Inclination | $i$ |
| Project semi-major axis | $x$ |
| Pulsar mass – in units of $M_\odot$ | $m_A$ |
| Companion mass – in units of $M_\odot$ | $m_B$ |
| Total Mass | $M$ |

*Table 4: The values of PK parameters. Courtesy (Kramer 2016).*

Our focus here it the PK parameters that are concerned with GR, and so the following equations are the ones used.

$$\dot{\omega} = 3\left(\frac{P_b}{2\pi}\right)^{-5/3} G^{2/3}M^{2/3}(1-e^2)^{-1}$$ Eq. 44

$$\gamma = e\left(\frac{P_b}{2\pi}\right)^{1/3} G^{2/3}M^{-1/3}m_B\left(1+\frac{m_2}{M}\right)$$ Eq. 45

$$\dot{P}_b = -\frac{192\pi}{5}\left(\frac{P_b}{2\pi}\right)^{-5/3} m_A m_B G^{5/3}M^{-1/3}$$

$$\left(1+\frac{73}{24}e^2+\frac{37}{96}e^4\right)(1-e^2)^{-7/2}$$ Eq. 46

$$s = x\left(\frac{P_b}{2\pi}\right)^{-2/3} G^{-1/3} M^{2/3} m_B^{-1}$$ Eq. 47

$$r = Gm_B$$ Eq. 48

The value $s = \sin i$, and the convention that $M_\odot/c^3 = 1$.

The PK parameter $\dot{\omega}$, is the advance of the periastron. Where $P$ is the orbital period, $e$ is the eccentricity; $m_p$ is the mass of the pulsar and $m_c$ is the mass of companion[55]. While the constant, $T_\odot = GM_\odot/c^3 = 4.925490947\ \mu$s (Kramer 2016).

$$\dot{\omega} = 3T_\odot^{2/3}\left(\frac{P}{2\pi}\right)^{-5/3}\frac{1}{1-e^2}(m_p + m_c)^{2/3}$$ Eq. 49

**Appendix 1.2.1: Newtonian Equation of motion**

Newtonian theory provides an equation of motion formula for a non-circular orbit as a function of the planet's angular momentum, which is shown in Equation 50 (Hobson et al. 2006). The angular momentum, per unit mass, of the orbiting planet, is $h$, and $u \equiv 1/r$. The angle at which the perihelion advances is $\phi$.

$$\frac{d^2u}{d\phi^2} + u = \frac{GM}{h^2}$$ Eq. 50

**Appendix 1.2.2: Alternative equation for the perihelion advance**

Other factors, as well as GR, need to be taken into account to resolve the residual accurately. The work of Will (2014) provides a new equation of the perihelion advance per orbit, Equation 51. This also includes elements of the parametrized post-Newtonian [PPN] framework, and the solar quadrupole moment[56], $J_2 = (2.2 \pm 0.1) \times 10^{-7}$ (Xu et al. 2011). The value of the PPN constants used are, $\gamma$ is the amount of curvature that is produced, per unit mass, and the ratio of inertial mass to gravitational mass, for astronomical objects that differ from unity is denoted by, $\beta$. (Williams et al. 1976; Nordtvedt Jr 1982; Dicke 1961).

$$\Delta\phi = 42.''98\left(\frac{1}{3}(2 + 2\gamma - \beta) + 3 \times 10^{-4}\frac{J_2}{10^{-7}}\right)$$ Eq. 51

[55] The companion may be another pulsar.

[56] The solar quadrupole moment, $J_2$, expresses the distortion of the solar potential, due to solar oblateness.

Other researchers, also take into account the cosmological constant[57] $\Lambda$, for an additional causal process of the advance of any perihelion (Arakida 2013; Islam 1983). However, the contribution by $\Lambda$ to the precession rate is too small to be detected with Mercury or other planets in the solar system (Kerr et al. 2003).

**Appendix 2.1.1: Alternative Notational Derivation of the Shapiro delay**

The trajectory of photons is a null geodesic and as such has no *proper time* associated with it. To mathematically define Shapiro delay, we need to consider the equation of motion, or equation of energy for any massless particle, grazing the Sun (Hobson et al. 2006). This is shown below in Equation 52.

$$\dot{r}^2 + \frac{h^2}{r^2}\left(1 - \frac{2\mu}{r}\right) = c^2k^2$$

Eq. 52

The value, $\dot{r}$ is the radial differential. Where, $k = \dot{t}(1-2\mu/r)$, while $\dot{t}$ is the time differential. The value, $h = r^2\phi$, and lastly, $\mu = 2GM/c^2$. The following integration leads to a calculation based on *coordinate time*. Looking at Figure 24, we can see the physical definition of $r$ and $r_0$.

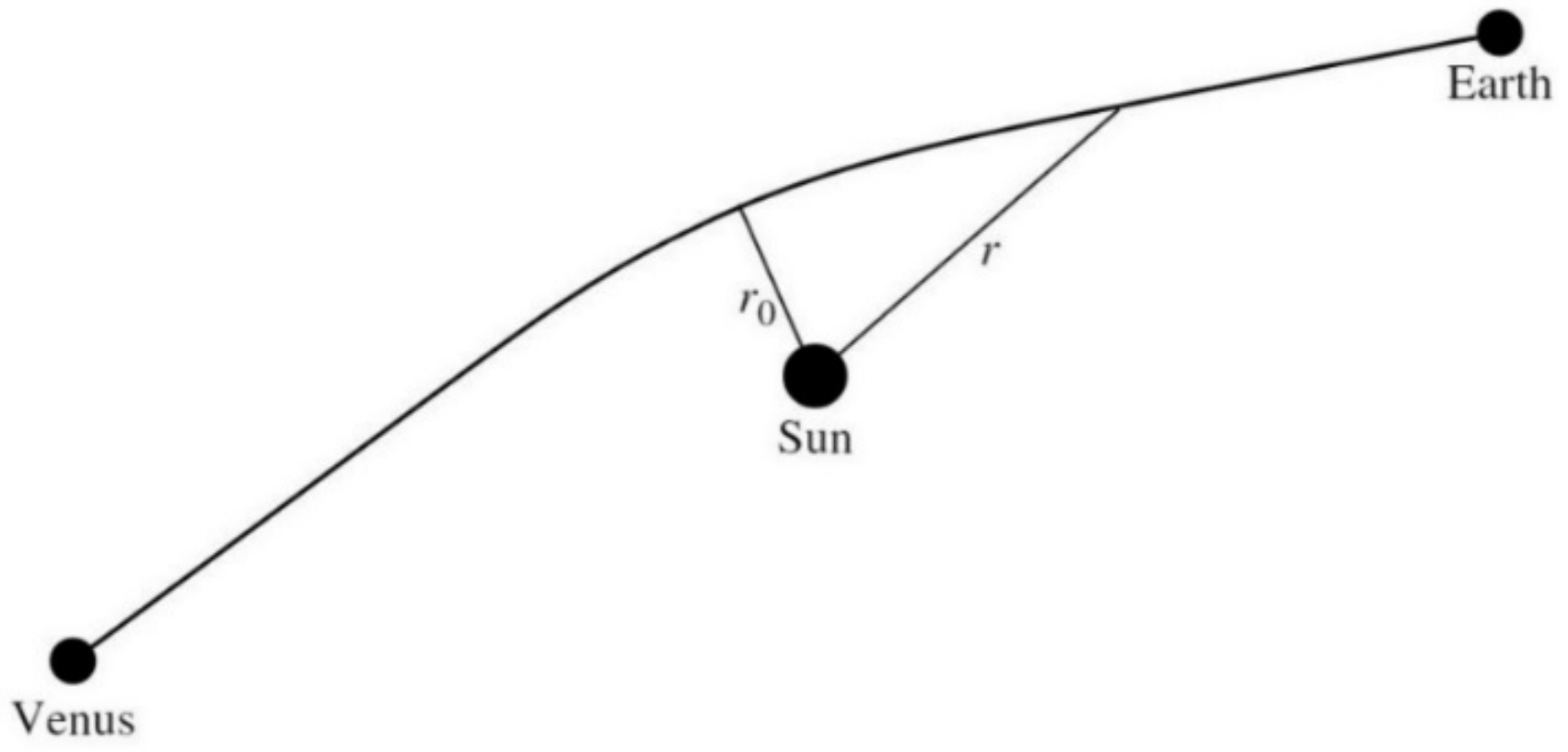

*Figure 24: radar reflection of photons from the Earth to Venus and back (Hobson et al. 2006).*

Hence, the integral between $r$ and $r_0$ can now be solved, as shown in Equation 53.

$$t(r, r_0) = \int_{r_0}^{r} \frac{1}{9(1-2\mu/r)}\left[1\frac{r_0^2(1-2\mu/r)}{r^2(1-2\mu/r_0)}\right]^{-1/2} dr$$

Eq. 53

The solution is the excess $\Delta t$, or Shapiro delay, in coordinate time, is provided in Equation 54. Note here, $r_e$ is the distance from the Earth to the Sun, and so too is, $r_v$ is the distance from Venus to the Sun (Hobson et al. 2006).

[57] $\Lambda$: the cosmological constant is the energy density of the vacuum of space.

$$\Delta t \approx \frac{4GM}{c^3}\left[\ln\left(\frac{r_e r_v}{r_0}\right)+1\right]$$ Eq. 54

**Appendix 2.2.1: Kerr Metric**

The Kerr metric is for spacetime near a massive rotating object and can be expressed by the line element, shown in Equation 55.

$$\mathrm{ds}^2 = \left(1-\tfrac{2m}{r}\right)\mathrm{d}t^2 - \left(1+\tfrac{2m}{r}\right)\left[\mathrm{d}r^2 + r^2 + r^2\left(\mathrm{d}\theta^2 + sin^2\theta d\phi^2\right)\right]$$
$$+\tfrac{4ma}{r}\sin^2\theta\ \mathrm{d}\phi\ \mathrm{d}t$$ Eq. 55

Equation 55 uses polar coordinates, r, $\theta$, and $\phi$. Furthermore, it is a linearised version of the Kerr metric (Rindler, 2006; Narlikar, 2010).

Moreover, a non-rotating BH that carries an electric charge is referred to as the Reissner−Nordstrom solution and it has the following form (Ryder 2009).

$$\mathrm{d}s^2 = -c^2\left(1-\frac{2m}{r}+\frac{q^2}{r^2}\right)\mathrm{d}t^2 + \left(1-\frac{2m}{r}+\frac{q^2}{r^2}\right)^{-1}\mathrm{d}r^2 + r^2\,\mathrm{d}\Omega^2$$ Eq. 56

And so:

$$q^2 = \frac{GQ^2}{4\pi\varepsilon_0 c^4},\quad Q = \text{electric charge}$$ Eq. 57

**Appendix 2.2.2: Other sources of errors**

Doppler shifts and diurnal motion[58] also need to be accounted for. Additionally, smaller relativistic effects need to be considered: changes to satellite clock frequencies by the Earth's quadrupole moment[59], differences between coordinate distances, relativistic invariant distances, and Shapiro delay (Ashby, 2002). There are minute effects, such as 'phase wrap-up'[60] and the gravitational effects of other solar system bodies, being most notably the Sun and the Moon (Ashby, 2003).

[58] Diurnal motion: daily motion of the celestial sphere across the sky. For the GPS, the correction is ~ $-1.2 \times 10^{-12}$.
[59] The Earth's quadrupole moment represents the varying gravitational potential due to the oblateness of the Earth.
[60] Phase wrap-up, is an effect of a signal's electric field vector rotating in opposite directions at the receiver.

The errors in the GPS are compounded, from the delays by ionospheric chromatic dispersion and the tropospheric refraction, receiver noise and resolution, multipath[61] and other user errors, which comprise ~ 95% of all errors (Grimes, 2008). There are values associated with other relativistic effects and caused by the rotation of the Earth and the tides.

### Appendix 2.4.1: Time-varying constant tests

***Time-Varying G***

A time-varying $G$ which is decreasing, would cause the lunar period and mean distance from the Earth to the moon to increase. This variation $\dot{G}/G$ is depicted in Equation 58, where $\dot{G}$ is the time varying $G$. Additionally, $n$ is the change in motion of the moon and $a$ is the lunar semi-major axis, as shown in Equation 58.

$$\dot{G}/G = 2\dot{n}/n + 3\dot{a}/a \qquad \text{Eq. 58}$$

Williams et al., (1996) determined the time variation of $G$ from LLR, was $\dot{G}/G \leq 8 \times 10^{-12}$ yr$^{-1}$. Then in 2004, this result was improved to $\dot{G}/G = (4 \pm 9) \times 10^{-13}$ yr$^{-1}$ (Williams et al., 2004).

More recently, the deviation of $G$, using data from the recombination epoch $G_{rec}$ and the present epoch $G$, was calculated to be in the order of, $G_{rec}/G < 1.0056$, whereby a non-varying $G = 1$ (Ooba et al., 2016).

***Variable Speed of Light [VSL]***

Einstein himself first mentioned the possibility of a variable speed of light [VSL] in 1907 (Einstein, 1907), and more comprehensively in 1911 (Einstein, 1911b). He concluded that $c$ is constant in a constant $\Phi$, but would vary in a varying $\Phi$. Einstein in the 1911 paper neglected the varying of fundamental length scale in different $\Phi$, he stated that "The principle of constancy of the speed of light can be upheld only when one restricts oneself to regions of constant gravitational potential," (Einstein, 1911b).

Dicke (1957) also expounded the idea of VSL as a theory of gravity, "The velocity of light in a 'bare' space could be greatly different from $c$ or even meaningless." According to Will, the idea of a varying $c$ does not fundamentally contradict general relativity. In fact, it is implicit in general relativity, as part of the description of coordinate space (Will, 1993). Moreover, the idea of VSL has re-emerged in the past few decades.

[61] Multipath is the propagation phenomenon from signals reaching the receiving device by two or more paths.

The Schwarzschild metric implies an anisotropy in $c$ with a varying $\Phi$. This anisotropy is given for mass $M$, at a distance $R$ from its centre by Equation 59.

$$\Delta c/c = GM/Rc^2 \equiv \gamma \qquad \text{Eq. 59}$$

There is a difference $\Delta c$, which should occur when the value $c$ is parallel to the equal gravitational potential lines. An upper limit was found: $\Delta c/c \sim 3 \times 10^{-11}$ (Shamir & Fox, 1969).

When measuring the speed of light variations from Gamma Ray Bursters [GRBs], it was also found that, the average intrinsic energy: $E_{\text{source}} = (1 + z)E_{\text{obs}}$.

Some *quintessence*[62] cosmology models suggest the accelerated expansion of the universe could be explained by the use of VSL (Chakraborty, 2002).

***Varying Fine Structure Constant $\alpha$***

Sommerfeld introduced the dimensionless fine structure constant α in 1916. It is referred to as the coupling constant, being a measure of the strength of the electromagnetic interaction.

The value of the α is small, $\alpha \cong 1/137$, while its current value is, $\alpha = 7.2974 \times 10^{-3}$ but it can be represented in different ways. Equation 60 is for $\alpha$ and is in SI units (Tobar, 2005).

$$\alpha = \frac{e^2}{4\pi e_0 \hbar c} \qquad \text{Eq. 60}$$

Where $h$ is Planck's constant and $\hbar = h/2\pi$. Additionally, $e$ is the electron charge, and $\varepsilon_0$ is the permittivity of free space.

The value of $\alpha$ seemed something of a mystery until the discovery of the quantum Hall effect [QHE] which independently corroborated the α from the electron magnetic moment. Finding a time varying value of $\alpha$ would violate the SEP.

See Table 5 for varying-constant constraints.

[62] Quintessence is a hypothetical concept of dark energy, which is a scalar field. It was postulated as an explanation of the observed accelerating expansion of the universe, as oppose to a true cosmological constant.

| **Constant k** | **Limit on $\dot{k}/k$ (yr$^{-1}$)** | **Redshift** | **Method** |
|---|---|---|---|
| Fine-structure constant | $< 1.3 \times 10^{-16}$ | 0 | Clock comparisons |
| $\alpha_{EM} = e^2/hc$ | $< 0.5 \times 10^{-16}$ | 0.15 | Oklo[63] Natural Nuclear reactor |
| | $< 3.4 \times 10^{-16}$ | 0.45 | $^{187}$Re decay in meteorites |
| | $(6.4 \pm 1.4) \times 10^{-16}$ | 0.2−3.7 | Spectra in distant quasars |
| | $< 1.2 \times 10^{-16}$ | 0.4−2.3 | Spectra in distant quasars |
| Weak interaction constant | $< 1 \times 10^{-11}$ | 0.15 | Oklo Natural Nuclear reactor |
| $\alpha_W = G_f m_p^2 c/h^3$ | $< 5 \times 10^{-12}$ | $10^9$ | Big-bang nucleosynthesis |
| e-p mass ratio | $< 3.3 \times 10^{-15}$ | 0 | Clock comparisons |
| | $< 3 \times 10^{-15}$ | 2.6−3.0 | Spectra in distant quasars |

*Table 5: Bounds on cosmological temporal variation of fundamental non-gravitational constants (Will, 2015).*

### Appendix 3.2.1: Polarised Gravitational Waves

Einstein's theory of gravity predicts distortion of GW with two degrees of polarization; alternative theories predict more polarizations (Lee, 2008). The GW curvature tensor ***h*** is considered a GW field. There is no *z* component, hence for GR there are only two possible polarisation states, these are represented by Equations 61 and 62 (Ju et al., 2000).

$$h_+ = h_{xx} = \mathrm{Re}\,[A_+ e^{-i\omega(t-z/c)}] \qquad \text{Eq. 61}$$

$$h_\times = h_{xy} = \mathrm{Re}\,[A_\times e^{-i\omega(t-z/c)}] \qquad \text{Eq. 62}$$

Where $A_+$ and $A_\times$ are the strains amplitudes of each polarisation state. Figure 25 shows the two polarisations in GR, graphically.

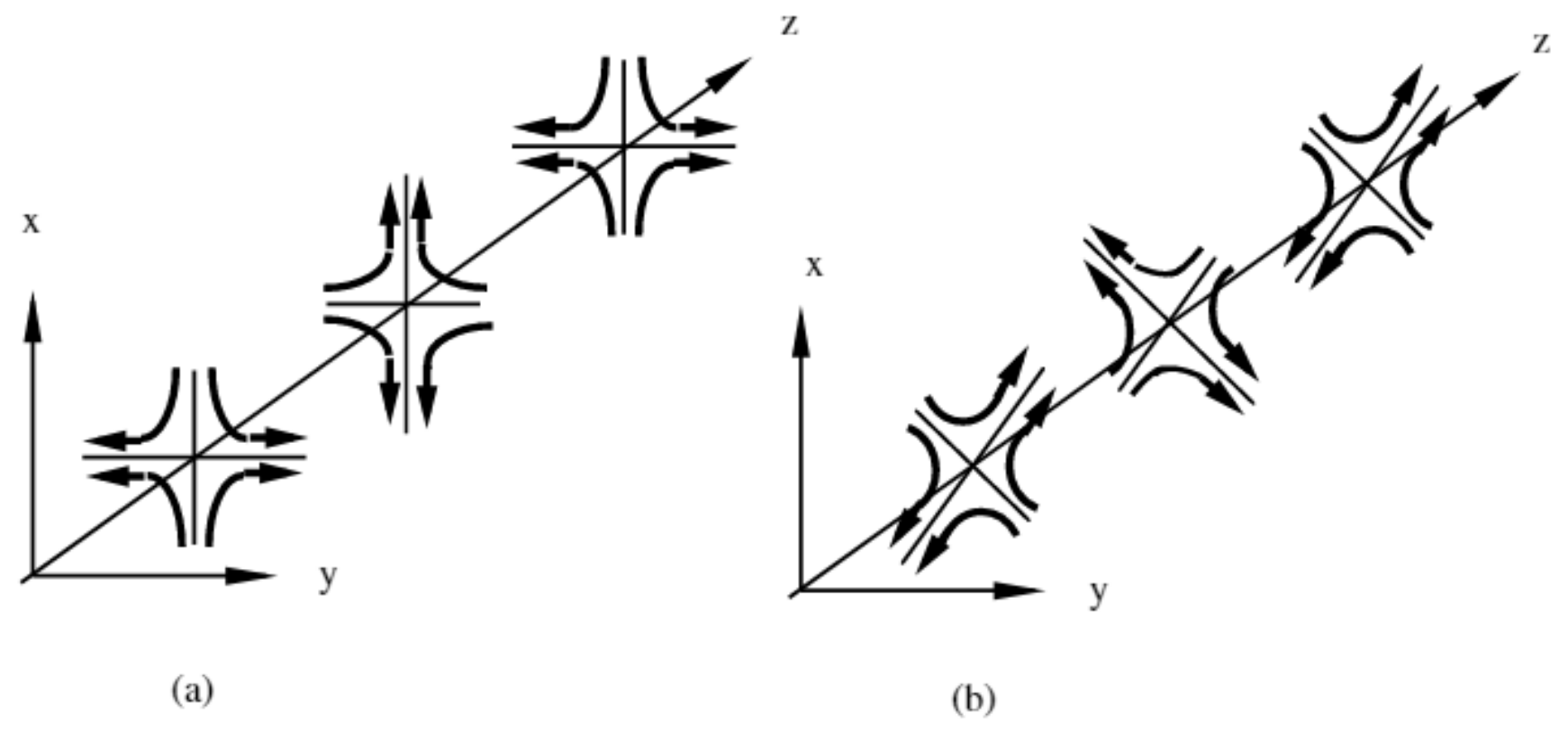

*Figure 25: GW field lines. ( a ) '+' polarisation; ( b ) '×' polarisation (Ju et al. 2000).*

[63] Oklo is a natural self-sustaining nuclear fission reactor whose fuel is a uranium deposit and is situation in Gabon.

**Appendix 4.3: Future missions and surveys**

Future missions will allow higher constraints to be placed on tests of GR and a few of these are briefly mentioned. Only aspects of GR testing will be considered.

***GAIA***

GAIA is a current astrometric space mission of ESA which was launched in December 2013 and is currently orbiting Lagrange L2. GAIA measures the proper motions and distances, using parallaxes, for more than one billion stars, down to ≅ 20 mag, at the sub-milliarcsecond level (Spagna et al., 2016).

As a result, GAIA's data will allow the determination of the PPN values $\gamma$ and $\beta$ within the solar system, from the deflection of light by Jupiter and other bodies.

GAIA's data will show any temporal variation in $\dot{G}/G$, and also the Nordtvedt parameter $\eta$ (Mouret, 2011), and monitor the precession of perihelion of eccentric NEOs (Bancelin et al., 2012). Moreover, the data will allow tests of the LLI, measurements of visual binary orbits, and also to estimate the energy flux of GWs (Klioner, 2014).

GAIA will provide an unprecedented determination of the $\gamma$ at the $10^{-6}$ level (Hestroffer et al., 2015). Although GAIA has been in operation for a few years, much of the data has still be analysed.

***Square Kilometre Array***

The Square Kilometre Array [SKA] is still under construction, in the deserts of Australia and South Africa, and is a project by the international SKA Organisation. The SKA will be a range of complex radio receivers which will work as an interferometer.

Essentially, the SKA will provide unprecedented accuracy in testing GR within the strong−field regime. In particular, timing measurements of binary pulsars, including BHs, to determine if any deviations of GR by way of orbital dynamics and the observed orbital decay parameter, $\dot{P}$. This will allow the estimation of a temporal variation of $\dot{G}/G$, the possible existence of dipole GWs and constrain the value of PK parameters.

The validity of LLI and LPI can also be ascertained as they are a consequence of deviations in GR orbital dynamics predictions of binary pulsars. Hence SEP can be verified, or invalidated, as a result. Additionally, the viability of the 'cosmic censorship conjecture'[64] and the 'no-hair theorem'[65] can be determined (Shao et al., 2014).

The growth rate of large-scale structure provides another test for deviations from GR, via redshift−space distortions[66] (Maartens et al., 2016). Table 6 shows the sensitivity to GWs for the SKA compared to other observatories.

| Detector | Frequency |
|---|---|
| Advanced LIGO | ~ 100 Hz |
| LISA | ~ mHz |
| SKA PTA | ~ nHz |

*Table 6: Frequency range limit of a SKA PTA in comparison to observatories detecting GW emission. Courtesy: (Kramer et al., 2004).*

***Euclid***

Euclid is an ESA mission, scheduled for launch 2018-2019, designed primarily to test dark energy or Λ, but it will also test the validity of GR on cosmic scales utilizing WL and measured BAO patterns, using a redshift survey of more than 50 million galaxies. Moreover, it will measure the growth index $\gamma_G$, with a precision better than ~0.02, compared to the GR value is $\gamma_G = 0.55$ (Majerotto et al., 2012).

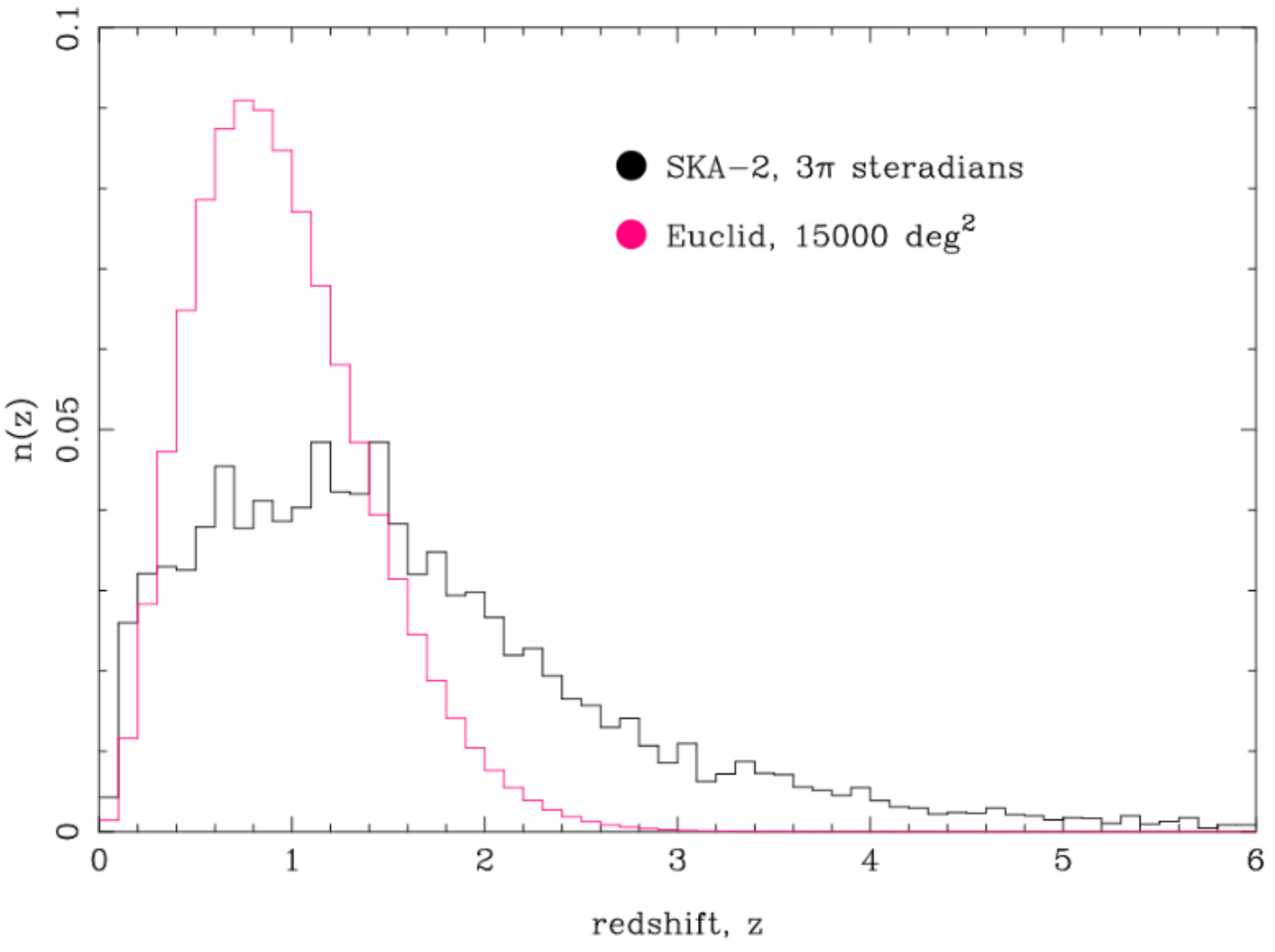

*Figure 26: Redshift distribution of sources against redshift for a 30,000 deg$^2$ WL survey with SKA2 compared to 15,000 deg$^2$ surveys with Euclid. Courtesy: (Maartens et al., 2016).*

[64] Penrose proposed the cosmic censorship conjecture: every BH singularity must have an event horizon.

[65] The no-hair theorem posits that all BH solutions of the Maxwell−Einstein equations of electromagnetism and gravitation in GR can be characterized by only, electric charge, mass and angular momentum.

[66] Redshift−space distortions are an observational effect, where the distribution of galaxies spatially appears distorted and squashed, when their positions are plotted a function of their redshift.

Survey comparisons with the SKA and Euclid are shown in Figure 26.

The Euclid data will provide constraints of the two relativistic potentials Ψ and Φ and provide a test of an indirect premise of GR: the cosmological principle (Amendola et al., 2016).

***LISA, eLISA and LISA Pathfinder***

The Laser Interferometer Space Antenna [LISA] was originally a planned space-borne GW detector by NASA and the ESA; then, it became eLISA and developed only by the ESA and is simply referred to a LISA. LISA Pathfinder[67] was launched in 2015, being a test mission for LISA. LISA Pathfinder has been successful, and ESA has started the proposal process. LISA will occupy Lagrange point L3 and consist of three arms using six active lasers, consisting of three spacecraft in a triangular shape and each being separated by $2.5 \times 10^9$ m (Amaro-Seoane et al., 2017).

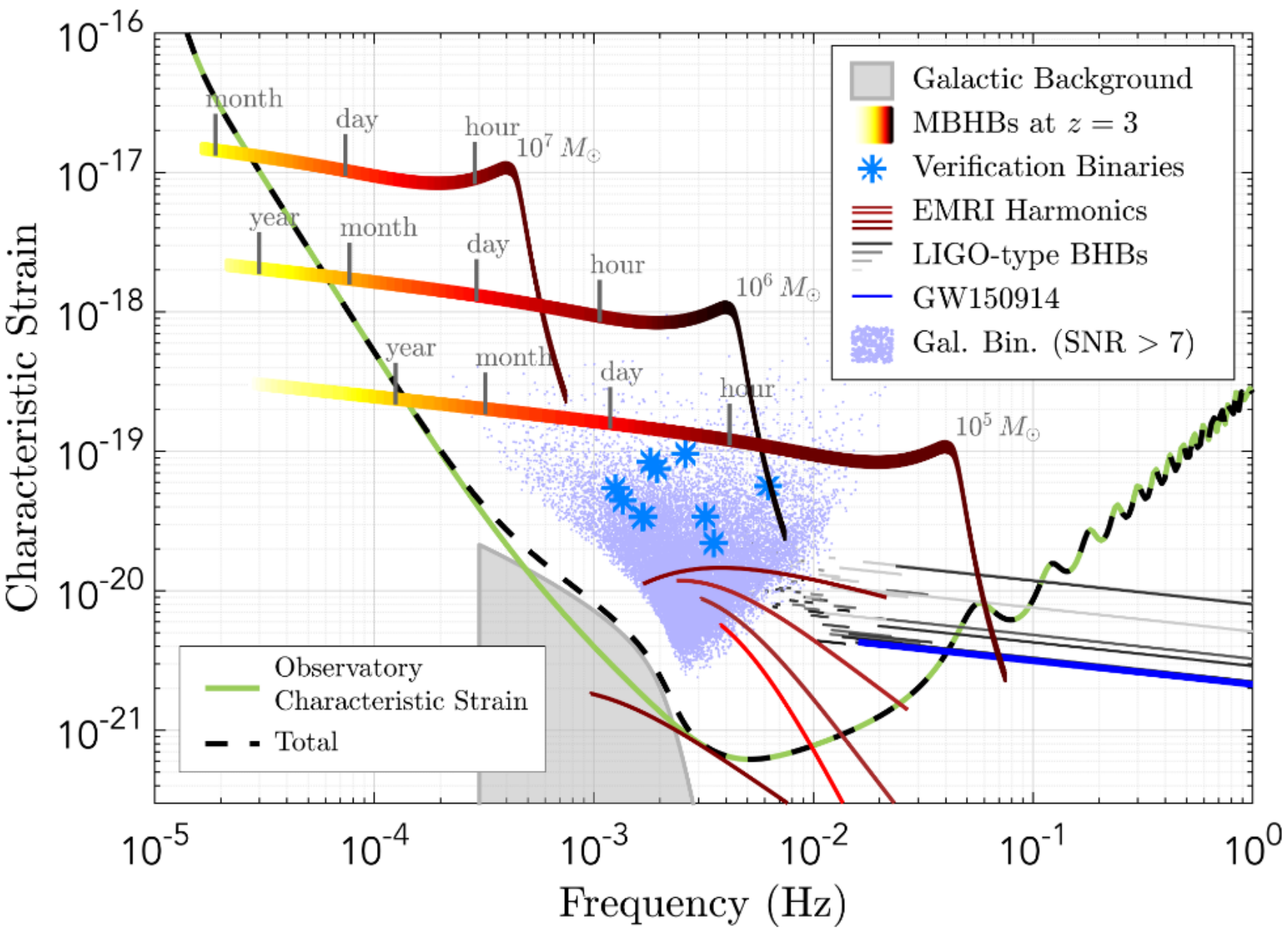

*Figure 27: Shows likely frequency range of LISA for GWs. The plot is 'characteristic strain amplitude' against the source frequency, showing massive black hole Binaries[MBHB] and black hole Binaries[BHB]. This data is at z = 3 with total masses of $M_\odot 10^7$, $M_\odot 10^6$ and $M_\odot 10^5$. (Amaro-Seoane et al., 2017).*

LISA will be able to determine the luminosity distance of massive HBs, with an uncertainty of ~10%, out to $z \cong 4$ for binaries of mass, $(10^6 + 10^6)\ M_\odot$ and out to $z \cong 2$ for binaries of mass, $(10^7 + 10^7)\ M_\odot$. Moreover, LISA will operate at the frequency waveband of $10^{-4}$ Hz and $10^{-1}$ Hz, allowing the detection of inspiralling stellar-mass compact neutron stars or BHs, with BH mass ranges of $\geq 10^2\ M_{\odot}$ (Berti et al., 2005). LISA will provide robust tests of GR after its launch in the 2030s, by way of sensitive GW detections. See Figure 27 for a chart of LISA's frequency range.

[67] LISA Pathfinder was originally called: Small Missions for Advanced Research in Technology-2 (SMART-2).

There are other missions and surveys that will be used for GR testing and that are not listed.